\documentclass[aps,pre,
showkeys,
longbibliography,
reprint,                %% preprint, draft, 
twocolumn,              %% onecolumn, 
nofootinbib,
floatfix]{revtex4-2}

\usepackage[T1]{fontenc}
\usepackage[utf8]{inputenc}
\usepackage{amsmath}
\usepackage{amsfonts}
\usepackage{graphicx}

\usepackage[labelformat=simple]{subcaption}

\usepackage{xcolor}
\definecolor{codegreen}{rgb}{0,0.6,0}
\definecolor{codegray}{rgb}{0.5,0.5,0.5}
\definecolor{codepurple}{rgb}{0.58,0,0.82}
\definecolor{backcolour}{rgb}{0.95,0.95,0.92}

\usepackage{listings}
\lstdefinestyle{mystyle}{
    backgroundcolor=\color{backcolour},   
    commentstyle=\color{codegreen},
    keywordstyle=\color{magenta},
    numberstyle=\tiny\color{codegray},
    stringstyle=\color{codepurple},
    basicstyle=\ttfamily\footnotesize, %% tiny
    breakatwhitespace=false,         
    breaklines=true,                 
    captionpos=t,                    
    keepspaces=true,                 
    numbers=left, numberstyle=\tiny, stepnumber=2, numbersep=5pt,
    frame=lines,
    showspaces=false,                
    showstringspaces=true,
    breakautoindent=false,
    showtabs=false,                  
    tabsize=2
}
\usepackage{cases}
\usepackage{lipsum}
\usepackage{soul}

\usepackage[colorlinks=true,hyperfootnotes=true,breaklinks=true,allcolors=teal]{hyperref}
\usepackage[capitalise]{cleveref}

\begin{document}

\title{Fine structure of phase diagram for social impact theory}

\author{Krzysztof Malarz}
\thanks{\href{https://orcid.org/0000-0001-9980-0363}{0000-0001-9980-0363}}
\affiliation{\href{https://ror.org/00bas1c41}{AGH University}, Faculty of Physics and Applied Computer Science, al.~Mickiewicza~30, 30-059 Krak\'ow, Poland}

\author{Maciej Wołoszyn}
\thanks{\href{https://orcid.org/0000-0001-9896-1018}{0000-0001-9896-1018} correspond. author}
\email{woloszyn@agh.edu.pl}
\affiliation{\href{https://ror.org/00bas1c41}{AGH University}, Faculty of Physics and Applied Computer Science, al.~Mickiewicza~30, 30-059 Krak\'ow, Poland}

\begin{abstract}
In this paper, the social impact theory introduced by Latan\'e is reconsidered.
A fully differentiated society is considered; that is, initially every actor has their own opinion.
The equivalent of Muller's ratchet guards that---even for the non-deterministic case (with a positive social temperature)---any opinion once removed from the opinion space does not appear again.
With computer simulation, we construct the phase diagram for Latan\'e model based on the number of surviving opinions after various evolution times.
The phase diagram is constructed on the two-dimensional plane of model control parameters responsible for the effective range of interaction among actors and the social temperature.
Introducing the Muller's ratchet-like mechanism gives a non-zero chance for any opinion to be removed from the system.
We believe that in such a case, for any positive temperature, ultimately a consensus is reached.
However, even for a moderate system size, the time to reach consensus is very long.
In contrast, for the deterministic case (without social temperature), the system may be frozen with clusters of actors having several different opinions, or even reach the cycle limit (with blinking structures).
\end{abstract}

\date{\today}


\maketitle

% CHAOS [ https://publishing.aip.org/resources/researchers/author-instructions/#CHA ]
% Lead Paragraph
% The first paragraph of the article should be the Lead Paragraph and contain the main points of the article, providing the “big picture” in a way that can be understood by non-specialist readers.

\textbf{%
Analysis of dynamics of social opinion is possible with many different models.
One of them is based on the theory of social impact, which predicts that individuals present to the society and disseminate their view on a given issue, and at the same time also pay attention to opinions promoted by others.
We use this approach to find out how difficult it is to reach a consensus state if the society is initially completely individualistic, with every actor presenting a unique opinion, not shared with any fellow member of the same society.
As an additional factor, we also introduce a random factor, which can play a role of `social temperature'.
Although in general reaching consensus is possible, it may take a very long time.
When the dynamics is purely deterministic the system can be `frozen' in a state of several clusters of opinions.
We show how many opinions survive depending on the observation time and what is the influence of `social temperature' and effective range of interaction among actors on this number of opinions.
Finally, we identify the set of model parameters where consensus is quite easily reached, where society polarization is the most probable outcome of the evolution, and where, even for a long time of system evolution, more than two opinions survive.
}

%% ############################################################
\section{Introduction}
%% ############################################################

The opinion dynamics remains a vivid part of sociophysics \cite{Sen_2014,Matjaz_2019,Jusup_2022,Zachary_2022}---the interdisciplinary branch of science that uses tools and methods of statistical physics to solve the problems with which sociologists fight in their everyday activities.

The models of opinion formation and dynamics \cite{Castellano_2009,Sobkowicz_2019,Hassani_2022} may be divided into two main groups with respect to the spectrum of opinions: with continuous (see, for instance, References~\onlinecite{Weisbuch_2002,Hegselmann-2002,Malarz2006b,Wu2022,Chica_2024}) or discrete opinions available in an artificial society.
Among the latter, the models most studied are: voter model \cite{Clifford_1973,Holley_1975,Liggett_1999,Liggett_2005}, majority-rule model \cite{Galam_2002}, Sznajd model \cite{Sznajd-2000} or models based on social impact \cite{Nowak-1990}.

The latter is based on Latan\'e theory of social impact \cite{Latane-1976,Latane-1981}.
By the way, the publication of Latan\'e paper \cite{Latane-1981} coincides with the birth of sociophysics, which is believed to be forty years old \cite{SI_70Galam_40Sociophysics}.
Latan\'e himself liked to think of this theory as `a light bulb theory of social relations' \cite{Latane-1981}, unintentionally making a contribution to the development of sociophysics. 
In this approach, every actor at the site $i$ in every discrete time step $t$ plays a role:
\begin{itemize}
    \item of a monochrome light source (the actor illuminates others in one of $K$ available colors $\Lambda_k$, i.e., shows and sends opinion $\lambda_i(t)=\Lambda_k$);
    \item and a full-spectrum light decoder (the actor detects which color $\Lambda_k$ gives the highest light illuminance at the site $i$).
\end{itemize}
Based on these observed illuminances (impacts), the actors can change their opinion in the subsequent time step $(t+1)$ to that which has the strongest illuminance (impact) on them.
Boltzmann-like factors yield probabilities of selecting $\Lambda_k$ as the opinion adopted by the actor in the $(t+1)$ time step in the non-deterministic version of algorithm \cite{1902.03454}, when the non-zero social temperature \cite{Micro-sociology,Macro-sociology} (information noise) is considered. 

The earlier computerized model applications deal with:
\begin{itemize}
\item observation of a phase transition from unanimity of opinions to disordered state \cite{Holyst-2000,Mansouri_2021};
\item impact of a strong leader \cite{Kacperski_1999,Kacperski-2000} (which may be introduced also in other models \cite{Shi_2019}) and social media influencers \cite{Rak_2018} on opinion dynamics;
\item simulation of language change \cite{Nettle_1999};
\item modeling individual vaccination decision making \cite{Xia_2013};
\item modeling bullying phenomenon in classrooms \cite{Tseng_2014};
\item impact of in-person closures on non-medical prescription opioid use among pupils \cite{Shojaati_2023}, {\it etc}.
\end{itemize} 

In addition to binary opinion models, discrete systems containing more than two opinions were previously investigated for the voter model \cite{Hadzibeganovic_2008,Vazquez_2004,Szolnoki_2004,Castello_2006,Mobilia_2011,Starnini_2012,Mobilia_2023,2405.05114}, 
the Sznajd model \cite{Rodrigues_2005,Kulakowski2010,Doniec_2022,2405.05114}, 
the majority-rule model \cite{Gekle-2005,Lima_2012,Galam_2013,Wu_2018,Zubillaga_2022} 
and other \cite{Vazquez_2003,Xiong_2017,Ozturk_2013,Martins_2020,Li_2022}.
The Latan\'e model was also enhanced in this direction, to account for several available opinions \cite{1902.03454,2002.05451,2010.15736,2211.04183,2405.05114}.
Finally, Malarz and Masłyk introduced an initially fully differentiated society \cite{Maslyk_2023} into the Latan\'e model, where the number of available opinions is comparable to the system size.

In Reference \cite{Maslyk_2023}, the preliminary shape of the phase diagram for the Latan\'e model was obtained in the $(\alpha,T)$ parameter plane \cite[see Figure~3]{Maslyk_2023}, where $\alpha$ is responsible for the effective range of interaction between actors, and social temperature $T$ measures the level of information noise.
However, the analyzed values of $\alpha$ ranged from 1 to 6 with step 1, and from 0.5 to 2.5 every 0.25 for the values of $T$.

In the current paper, we return to this problem with a much more systematic approach in scanning both the model parameter responsible for the level of social noise ($T$) and the effective range of interactions ($\alpha$).
The construction of the phase diagram is based solely on the number of ultimately observed (surviving) opinions. 
The initial number of opinions is exactly equal to the number of actors; in other words, initially every actor has their own unique opinion.

We note, however, that the meaning of the `phase diagram' term in the paper title is a rather attractive marketing hook---as even defining the ordering parameter here is rather hard task \cite{Latane_Nowak_1994}.
In our opinion, the system governed by social impact theory tends ultimately to the consensus, (un)fortunately the time of reaching this consensus is extremely large even for relatively not too large system sizes.

%% ############################################################
\section{Model}
%% ############################################################

We adopt the original formulation of the computerized version of the social impact model proposed by Nowak {\it et~al}. \cite{Nowak-1990} after its modification \cite{1902.03454,2002.05451,2211.04183} to allow for a multitude of opinions $\Lambda_k$ and $k=1,\cdots,K$.
The opinion of the actor $i$ at time $t$ is $\lambda_i(t)$.
We assume $L^2$ actors that occupy nodes of the square lattice.
Every actor $i$ is characterized by two parameters:
\begin{itemize}
    \item supportiveness $s_i\in[0,1]$, which describes the intensity of interaction with actors currently sharing opinion $\lambda_i$,
    \item and persuasiveness $p_i\in[0,1]$---describing the intensity of interaction with believers of different opinions than currently adopted by the actor $i$.
\end{itemize}
The supportiveness $s_i$ and the persuasiveness $p_i$ parameters are equivalents of the powers of the light bulbs in terms of ‘a light bulb theory of social relations’ \cite{Latane-1981}.

The social impact exerted on the actor $i$ by the actors $j=1,\ldots,L^2$, sharing the opinion $\Lambda_k$, is
%% ================================================================
\begin{subequations}
\label{eq:szamrej}
%% ----------------------------------------------------------------
\begin{align}
& \mathcal{I}_{i;k}(t) = \nonumber\\
    &\quad \sum_{j=1}^{L^2}{\frac{4s_j}{g(d_{i,j})} \cdot \delta(\Lambda_k, \lambda_j(t)) \cdot \delta(\lambda_j(t),\lambda_i(t))} \label{eq:szamrej_sum_same}\\
    &+ \sum_{j=1}^{L^2}{\frac{4p_j}{g(d_{i,j})} \cdot \delta(\Lambda_k, \lambda_j(t)) \cdot [1-\delta(\lambda_j(t),\lambda_i(t))}], \label{eq:szamrej_sum_diff}
\end{align}
\end{subequations}
%% ================================================================
where $g(\cdot)$ is an arbitrarily chosen function that scales the Euclidean distance $d_{i,j}$ between actors $i$ and $j$, and the Kronecker delta $\delta(x,y)=0$ when $x\ne y$ and $\delta(x,y)=1$ when $x=y$.
The combination of Kronecker deltas prevents the occurrence of terms describing the interaction between actors with different opinions in the summation \eqref{eq:szamrej_sum_same}.
Thus, we have such terms only if $\lambda_i(t)=\lambda_j(t)$, and we use actors' supportiveness $s_j$ to calculate the social impact. 
It also prevents the appearance of terms that describe interaction between actors with the same opinions \eqref{eq:szamrej_sum_diff} and thus we have nonzero terms only if $\lambda_i(t)\ne\lambda_j(t)$ and we use actors' persuasiveness $p_j$ to calculate the social impact.

According to the social impact theory \cite{Latane-1981}, the impact of the more distant actors should be smaller than that of the closest ones.
Thus, the $g(\cdot)$ function should be an increasing function of its argument.
Here, we assume that
%% ----------------------------------------------------------------
\begin{equation}
\label{eq:fg}
g(x) = 1+x^\alpha,
\end{equation}
%% ----------------------------------------------------------------
where the exponent $\alpha$ is a model control parameter while the first additive component ensures finite self-supportiveness.

Dworak and Malarz showed that for $\alpha = 2$, roughly 25\% of the impact comes from nine nearest neighbors (when the investigated actor occupies the center of a $3\times 3$ square).
This ratio increases to $\approx 59\%$, $\approx 80\%$ and $\approx 96\%$ for $\alpha = 3$, 4 and 6.
Calculating the relative impact exerted by actors from the neighborhood reduced to $5\times 5$ square gives roughly 39\%, 76\%, 92\%, and 99\% of the total social impact for $\alpha = 2$, 3, 4, and 6, respectively \cite[see Ref.][pp. 5--6, Fig. 2, Tab. 1]{2211.04183}. 
Dworak and Malarz concluded that ``the $\alpha$ parameter says how influential the nearest neighbors are with respect to the entire population: the larger $\alpha$, the more influential the nearest neighbors are''.

Here, we decided to use random values of $p_i$ and $s_i$.
Initially, at $t=0$, each actor has their own unique opinion $\lambda_i(t=0)=\Lambda_i$.

In the deterministic version of the algorithm, social impacts \eqref{eq:szamrej} yield the opinion of the actor $i$ in time $(t+1)$,
%% ----------------------------------------------------------------
\begin{equation}
\begin{split}
\label{eq:Teq0}
\lambda_i(t+1)=\Lambda_k\\ 
\iff \mathcal I_{i,k}(t)=\max(\mathcal I_{i,1}(t), \mathcal I_{i,2}(t),\cdots,\mathcal I_{i,K}(t)),
\end{split}
\end{equation}
%% ----------------------------------------------------------------
In other words, the actor $i$ adopts the opinion that exerts the largest social impact on them.

In the probabilistic version of the algorithm, the social impacts \eqref{eq:szamrej} imply Boltzmann-like\footnote{Please note, that adding minus sign before summation signs in \Cref{eq:szamrej}, minus sign in Equation~\eqref{eq:prob_Igt0} before $\mathcal{I}_{i,k}$, and the change of the function $\max(\cdot)$ to $\min(\cdot)$ in \Cref{eq:Teq0} provides exact Boltzmann factors, but the description of the deterministic case also requires the change of narration from maximal impact, either to the maximum absolute value of impact or to the lowest impact. In the latter case, the social impact \eqref{eq:szamrej} starts to mimic the system energy.} probabilities
%% ----------------------------------------------------------------
\begin{subnumcases}{\label{eq:probability_p}p_{i,k}(t)=}
    0                            & $\iff \mathcal{I}_{i,k}=0$, \label{eq:prob_Ieq0}\\
    \exp(\mathcal{I}_{i,k}(t)/T) & $\iff \mathcal{I}_{i,k}>0$, \label{eq:prob_Igt0}
\end{subnumcases}
%% ----------------------------------------------------------------
that the actor $i$ adopts the opinion $\Lambda_k$.
The parameter $T$ plays a role of social temperature \cite{Micro-sociology,Macro-sociology}.
Similarly to earlier approaches \cite{Maslyk_2023,2405.05114}, opinions with zero impact cannot be adopted by any of the actors.
In other words, according to \Cref{eq:prob_Ieq0}, the opinions with zero impact are not available.

Probabilities \eqref{eq:probability_p} require proper normalization ensured by
%% ----------------------------------------------------------------
    \begin{equation}
    \label{eq:probability_P}
    P_{i,k}(t) = \frac{p_{i,k}(t)}{\sum^K_{j=1} p_{i,j}(t)}.
    \end{equation}    
%% ----------------------------------------------------------------
Then, the time evolution of the opinion of actor $i$ is
%% ----------------------------------------------------------------
\begin{equation}
\label{eq:Tgt0}
\lambda_i(t+1)=\Lambda_k \text{ with the probability } P_{i,k}(t).
\end{equation}
%% ----------------------------------------------------------------

An example of deterministic evolution for a small system (with $L^2=9$ actors and $K=3$ opinions) and exact calculations of social impacts are given in \Cref{app:small}.

%% ############################################################
\section{Computations}
%% ############################################################

We implement \Cref{eq:szamrej,eq:fg,eq:Teq0,eq:probability_p,eq:probability_P,eq:Tgt0} as a Fortran95 code (see \Cref{lst:code} in \Cref{app:listing}).
In a single Monte Carlo step (MCS), every actor has a chance to change their opinion according to \Cref{eq:Teq0} or \Cref{eq:Tgt0}.
The system evolution takes $t_\text{max}$ MCS. 
The update of actors' opinions is performed synchronously.
The results are averaged over $R$ independent simulations, which allows for relatively easy parallelization of the computations, for example on multiple cores of the used CPU.

\begin{figure}
\includegraphics[width=.48\textwidth]{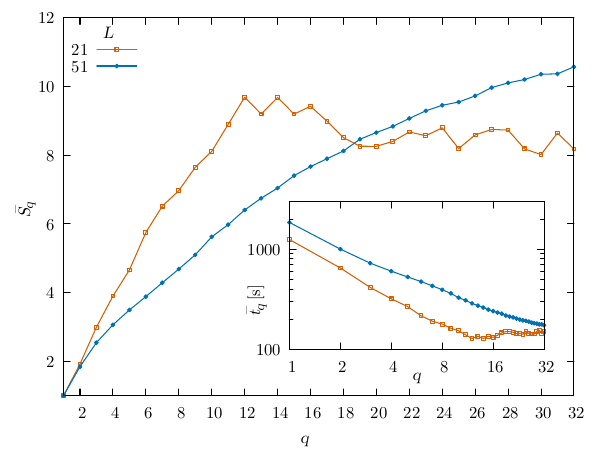}
\caption{\label{fig:speedup}
Average speedup $\bar{S}_q$ of the used code running in parallel on $q$ cores for $L=21$ with $t_\textrm{max}=10^4$ and $L=51$ with $t_\textrm{max}=10$; calculations consisting of $R=1680$ simulations repeated three times to obtain the average value.
The inset shows the corresponding average wall-clock times $\bar{t}_q$ in a log-log plot}
\end{figure}

\Cref{fig:speedup} shows the parallel speedup and calculation time for two test cases ($L=21$ and $L=51$) executed on a 32-core Intel Xeon(R) Platinum 8562Y+ CPU, with $t_\textrm{max}$ values chosen to have similar sequential times.
The speedup $S_q$ for calculations running in parallel on $q$ cores is $S_q = t_q / t_s$, where $t_s$ is the sequential time (one-core calculation), and $t_q$ is the time of the same calculation executed on $q$ cores.
For each system size, measurements were repeated three times and the averages $\bar{S}_q$ and $\bar{t}_q$ are presented in \Cref{fig:speedup}.

For $L=21$, the speedup is initially almost ideal, which results from the fact that all data from all threads can be stored in the cache of the processor, without the need to copy it from RAM. However, this is true only when the number of threads and used cores is below ca.~12.
In the case of $L=51$, even a single thread requires a large amount of memory, beyond the available cache of the CPU, which results in the speedup characteristics typical for Amdahl's law \cite{Amdahl_1967}.

%% ============================================================
%% ============================================================

%% ############################################################
\section{Results}
%% ############################################################

The simulations are carried out on a square lattice with open boundary conditions and with size $L=21$, that is, for an artificial society of 441 actors.
We assume random values of $s_i$ and $p_i$ taken uniformly from the interval $[0,1]$.

\begin{figure*}[htbp]
%% ------------------------------------------------------------
\begin{subfigure}[b]{0.32\textwidth}
\caption{\label{fig:phase_diagram-t1e3}}
\includegraphics[width=.99\textwidth]{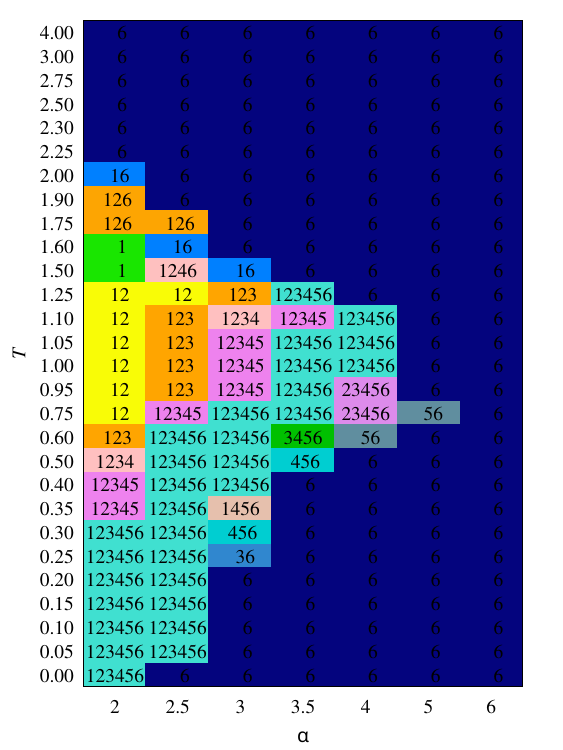}
\end{subfigure}
\hfill %% ------------------------------------------------------------
\begin{subfigure}[b]{0.32\textwidth}
\caption{\label{fig:phase_diagram-t1e5}}
\includegraphics[width=.99\textwidth]{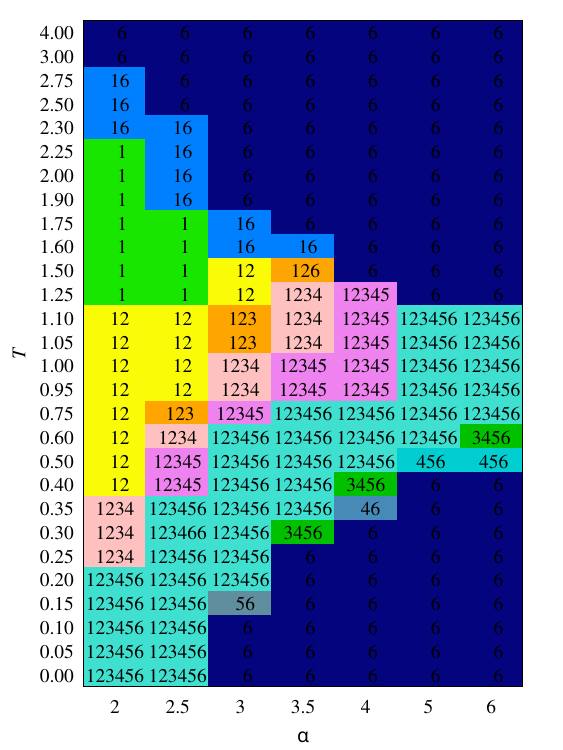}
\end{subfigure}
\hfill %% ------------------------------------------------------------
\begin{subfigure}[b]{0.32\textwidth}
\caption{\label{fig:phase_diagram-t1e6}}
\includegraphics[width=.99\textwidth]{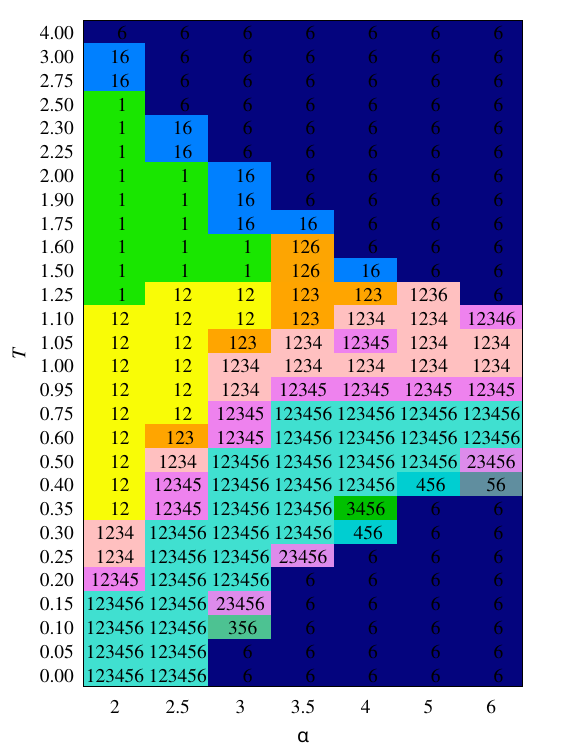}
\end{subfigure}
%% ------------------------------------------------------------
\caption{\label{fig:phase_diagram}Phase diagram of Latan\'e model showing the number of surviving opinions as observed in $R=10^3$ simulations, as a function of the social temperature $T$ and the $\alpha$ parameter related to the impact of the nearest neighbors relative to the impact of entire population. The numbers of opinions are encoded as follows:
`1' = unanimity;
`2' = polarization;
`3' = three opinions;
`4' = four opinions;
`5' = five opinions;
`6' = more than five opinions;
`12' = co-existence of phases 1 and 2;
`16' = co-existence of phases 1 and 6;
`1234' = co-existence of phases 1, 2, 3, 4;
`12345' = co-existence of phases 1, 2, 3, 4, 5;
`123456' = co-existence of phases 1, 2, 3, 4, 5, 6, {\it etc}.
In subsequent diagrams results after 
\subref{fig:phase_diagram-t1e3} $t_\text{max}=10^3$, 
\subref{fig:phase_diagram-t1e5} $t_\text{max}=10^5$,
\subref{fig:phase_diagram-t1e6} $t_\text{max}=10^6$ 
MCS are presented}
\end{figure*}

In \Cref{fig:phase_diagram} the phase diagram of the Latan\'e model is presented in the $(\alpha,T)$ parameter plane.
The different colors of `bricks' and different numerical sequences on them correspond to different final (that is, at the time $t=t_\text{max}$) states of the system observed in simulation.
The presence of number `1' in a sequence informs on the possibility of observation of opinion unanimity; `2'---on system polarization; `3', `4' and `5'---on 
%% ----------------------------------------------------------------
\begin{equation}
\label{eq:nou}
n_\text{o}^\text{u}\equiv n_\text{o}(t\to\infty)
\end{equation}
%% ----------------------------------------------------------------
equal to 3, 4, and 5, respectively.
The `6' indicates that finally more than five opinions were observed ($n_\text{o}^\text{u}>5$).
The mixture of labels, for instance `12', indicates co-existence of phases `1' and `2', `1234', indicates co-existence of phases `1',  `2',  `3' and `4', etc.
The subsequent diagrams show the evolution of the system after
$t_\text{max}=10^3$ [\Cref{fig:phase_diagram-t1e3}], 
$t_\text{max}=10^5$ [\Cref{fig:phase_diagram-t1e5}]
and $t_\text{max}=10^6$ [\Cref{fig:phase_diagram-t1e6}]  
MCS.

%% ============================================================
%% ============================================================

\begin{figure*}[htbp]
%% ------------------------------------------------------------
\begin{subfigure}[b]{0.32\textwidth}
\caption{\label{fig:max_no-t1e3}}
\includegraphics[width=.99\textwidth]{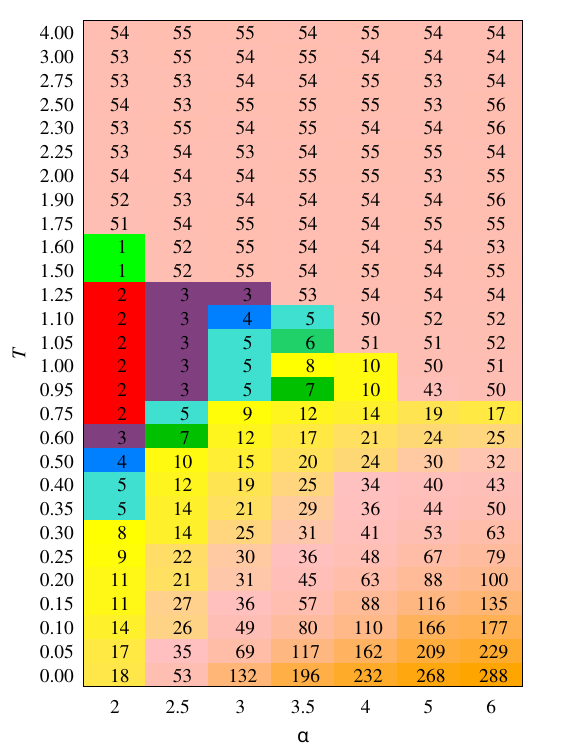}
\end{subfigure}
\hfill %% ------------------------------------------------------------
\begin{subfigure}[b]{0.32\textwidth}
\caption{\label{fig:max_no-t1e5}}
\includegraphics[width=.99\textwidth]{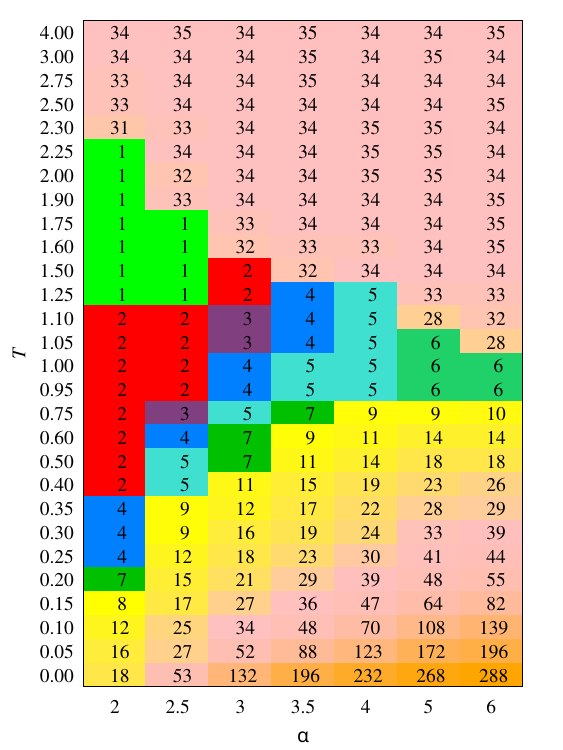}
\end{subfigure}
\hfill %% ------------------------------------------------------------
\begin{subfigure}[b]{0.32\textwidth}
\caption{\label{fig:max_no-t1e6}}
\includegraphics[width=.99\textwidth]{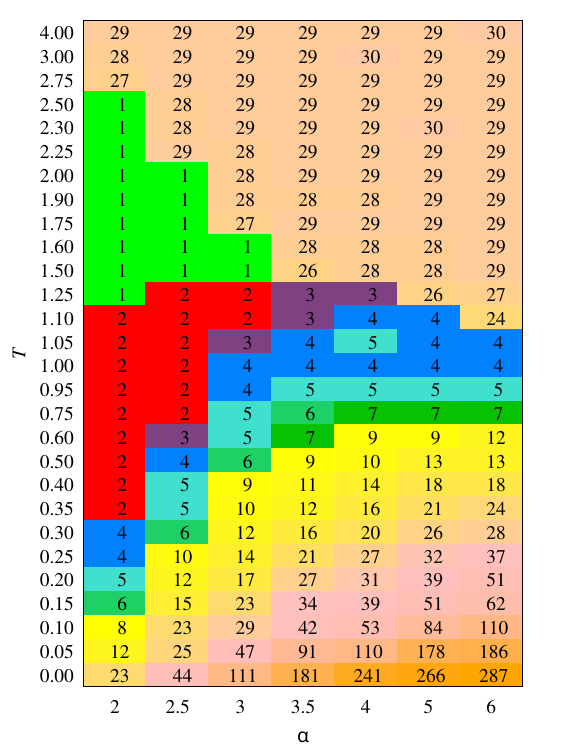}
\end{subfigure}
%% ------------------------------------------------------------
\caption{\label{fig:max_no}The largest number of surviving opinions $\max(n^\text{u}_\text{o})$ observed in $R=10^3$ simulations after 
\subref{fig:max_no-t1e3} $t_\textrm{max}=10^3$, 
\subref{fig:max_no-t1e5} $t_\textrm{max}=10^5$,
\subref{fig:max_no-t1e6} $t_\textrm{max}=10^6$ MCS.
The results are presented for varying social temperature $T$ and parameter $\alpha$ affecting to the effective range of interaction between actors, with $\max(n^\text{u}_\text{o})=1$ (light green background) corresponding to all simulations leading to consensus
}
\end{figure*}

In \Cref{fig:max_no} the largest numbers $\max(n^\text{u}_\text{o})$ of surviving opinions after
$t_\textrm{max}=10^3$ [\Cref{fig:max_no-t1e3}], 
$t_\textrm{max}=10^5$ [\Cref{fig:max_no-t1e5}],
$t_\textrm{max}=10^6$ [\Cref{fig:max_no-t1e6}] are presented. 
This allows us to distinguish various system behaviors and provide more detailed information on the final state of the system when the label `6' is indicated in the phase diagram given in \Cref{fig:phase_diagram}. 

%% ============================================================
%% ============================================================

\begin{figure*}[htbp]
%% ------------------------------------------------------------
\begin{subfigure}[b]{0.32\textwidth}
\caption{\label{fig:f_no_1}$n_\text{o}^\textrm{u}=1$}
\includegraphics[width=.99\textwidth]{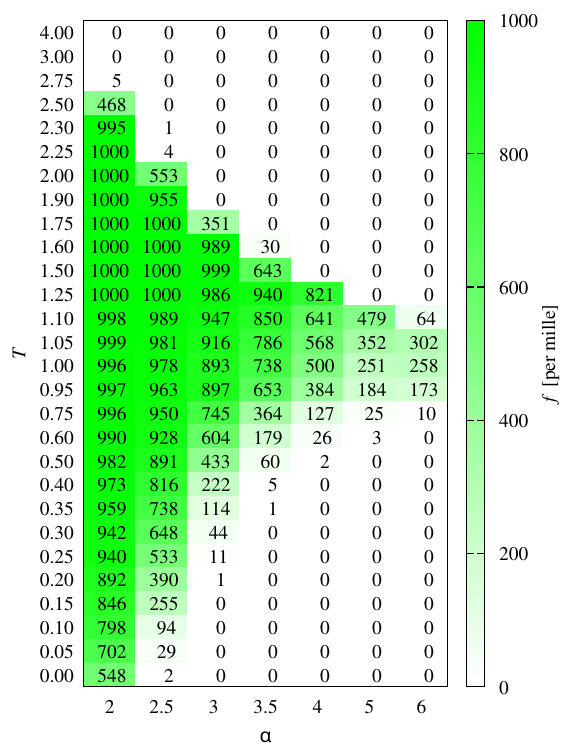}
\end{subfigure}
%% ------------------------------------------------------------
\begin{subfigure}[b]{0.32\textwidth}
\caption{\label{fig:f_no_2}$n_\text{o}^\textrm{u}=2$}
\includegraphics[width=.99\textwidth]{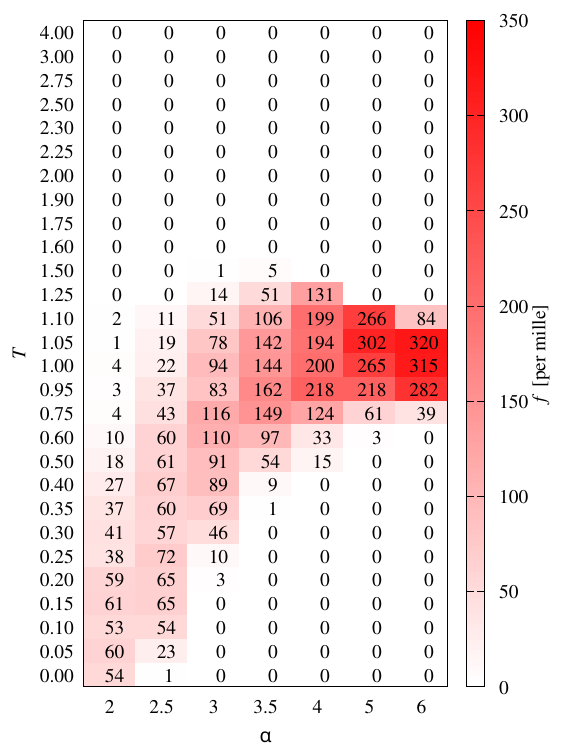}
\end{subfigure}
%% ------------------------------------------------------------
\begin{subfigure}[b]{0.32\textwidth}
\caption{\label{fig:f_no_3}$n_\text{o}^\textrm{u}=3$}
\includegraphics[width=.99\textwidth]{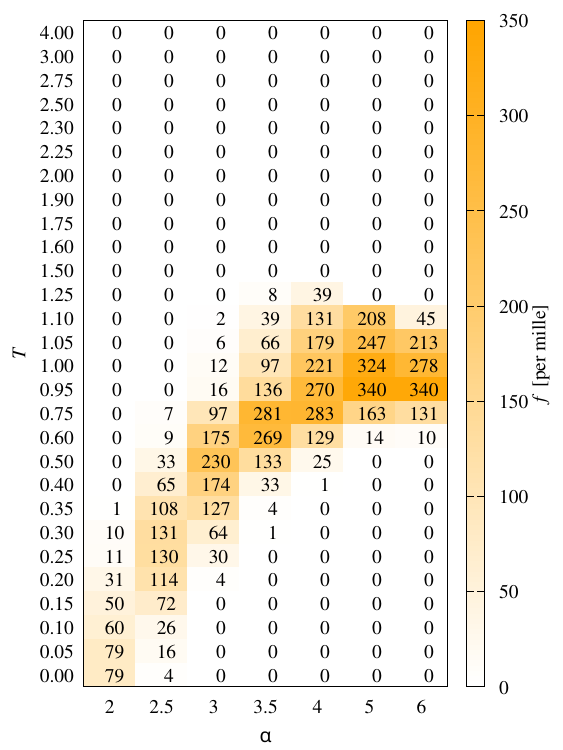}
\end{subfigure}
%% ------------------------------------------------------------
\begin{subfigure}[b]{0.32\textwidth}
\caption{\label{fig:f_no_4}$n_\text{o}^\textrm{u}=4$}
\includegraphics[width=.99\textwidth]{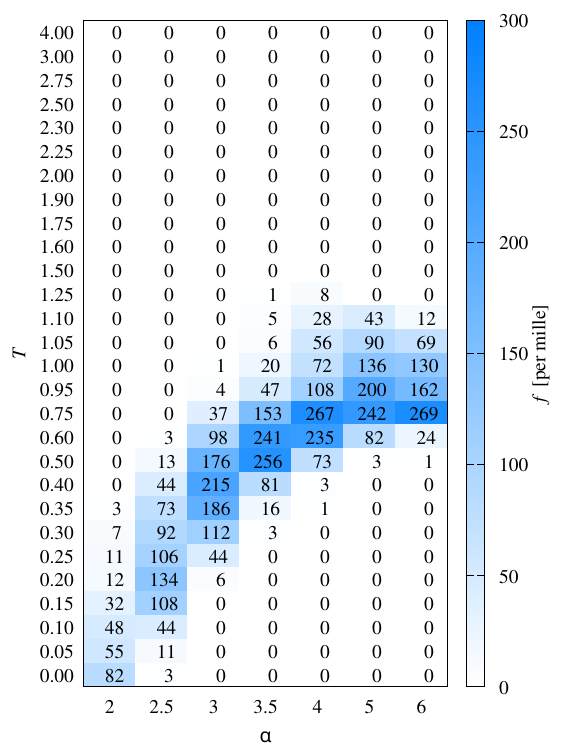}
\end{subfigure}
%% ------------------------------------------------------------
\begin{subfigure}[b]{0.32\textwidth}
\caption{\label{fig:f_no_5}$n_\text{o}^\textrm{u}=5$}
\includegraphics[width=.99\textwidth]{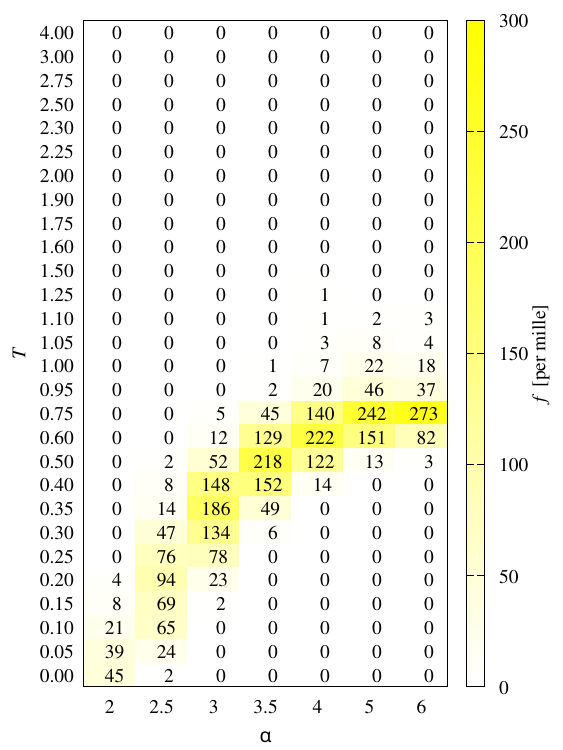}
\end{subfigure}
%% ------------------------------------------------------------
\begin{subfigure}[b]{0.32\textwidth}
\caption{\label{fig:f_no_6}$n_\text{o}^\textrm{u}>5$}
\includegraphics[width=.99\textwidth]{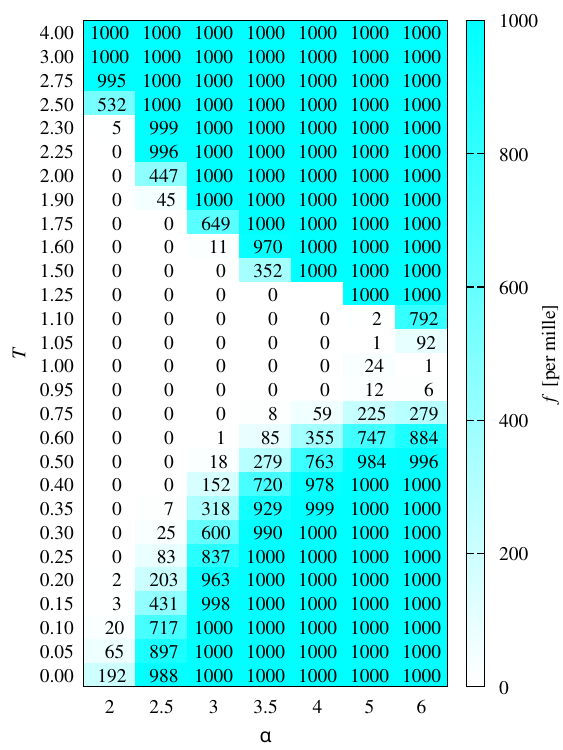}
\end{subfigure}
%% ------------------------------------------------------------
\caption{\label{fig:f}Frequency $f$ (in per mille) of ultimately surviving $n_\textrm{o}^\textrm{u}$ opinions in $R=10^3$ simulations. It shows how likely it is to have \subref{fig:f_no_1} $n_\textrm{o}^\textrm{u}=1$, \subref{fig:f_no_2} 2, \subref{fig:f_no_3} 3, \subref{fig:f_no_4} 4, \subref{fig:f_no_5} 5 and \subref{fig:f_no_6} more than 5 opinions after completing $t_{\textrm{max}}=10^5$ MCS for given values of the social temperature $T$ and parameter $\alpha$}
\end{figure*}

\Cref{fig:f} shows frequency $f$ (in per mille) of ultimately surviving $n_\textrm{o}^\textrm{u}$ opinions obtained in $R=10^3$ simulations after performing $t_{\textrm{max}}=10^5$ MCS. 
The subsequent figures indicate frequencies for
$n_\textrm{o}^\textrm{u}=1$ [\Cref{fig:f_no_1}], 
$n_\textrm{o}^\textrm{u}=2$ [\Cref{fig:f_no_2}], 
$n_\textrm{o}^\textrm{u}=3$ [\Cref{fig:f_no_3}], 
$n_\textrm{o}^\textrm{u}=4$ [\Cref{fig:f_no_4}], 
$n_\textrm{o}^\textrm{u}=5$ [\Cref{fig:f_no_5}] 
and 
$n_\textrm{o}^\textrm{u}>5$ [\Cref{fig:f_no_6}]. 

%% ============================================================
%% ============================================================

The detailed distributions of $n_\textrm{o}^\text{u}$ as functions of the social temperature $T$ for various parameters $\alpha$ after $t_\textrm{max}=10^3$ (see \Cref{fig:no_hist_tmax1e3}), $t_\textrm{max}=10^5$ (see \Cref{fig:no_hist_tmax1e5}) and $t_\textrm{max}=10^6$ (see \Cref{fig:no_hist_tmax1e6}) are presented in \Cref{app:distributions}.

%% ============================================================
\begin{figure}[htbp]
%% ------------------------------------------------------------
\begin{subfigure}[b]{0.48\textwidth}
\caption{\label{fig:tau_alpha2}$\alpha=2$}
\includegraphics[width=.99\textwidth]{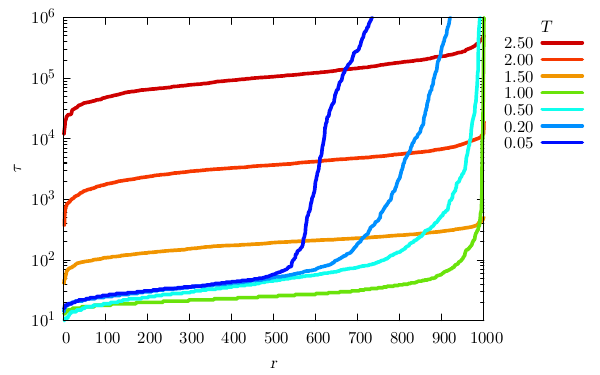}
\end{subfigure}
%% ------------------------------------------------------------
\begin{subfigure}[b]{0.48\textwidth}
\caption{\label{fig:tau_alpha3}$\alpha=3$}
\includegraphics[width=.99\textwidth]{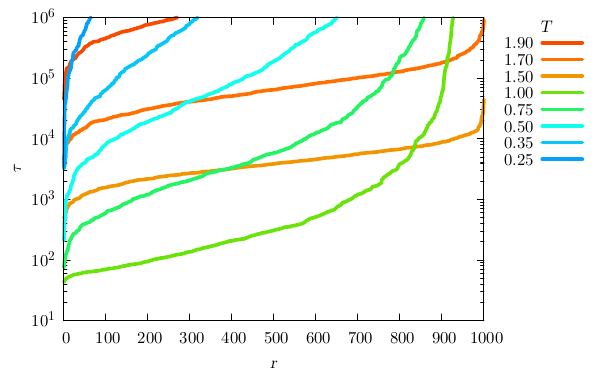}
\end{subfigure}
%% ------------------------------------------------------------
\begin{subfigure}[b]{0.48\textwidth}
\caption{\label{fig:tau_alpha4}$\alpha=4$}
\includegraphics[width=.99\textwidth]{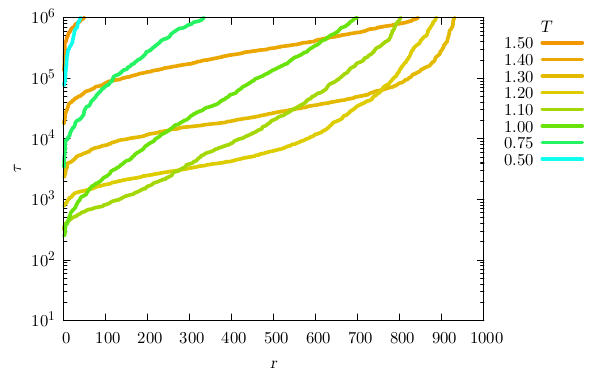}
\end{subfigure}
%% ------------------------------------------------------------
\caption{\label{fig:tau}Examples of the time $\tau$ of reaching the consensus ($n_\text{o}^\text{u}=1$) as dependent on the number $r$ of the performed simulation (ranked in ascending order) for \subref{fig:tau_alpha2} $\alpha=2$, \subref{fig:tau_alpha3} $\alpha=3$, \subref{fig:tau_alpha4} $\alpha=4$ and various temperatures $T$ after $t_\textrm{max}=10^6$}
\end{figure}
%% ============================================================

%% ############################################################
\section{\label{sec:discussion}Discussion}
%% ############################################################

In \Cref{fig:phase_diagram} we can observe the time evolution of the phase diagram for the social impact model.
During this evolution, subsequently the area covered with bricks labeled with `1' increases while area covered with bricks labeled `6' decreases.
This tendency is also reflected in \Cref{fig:max_no}, as the area covered by bricks labeled `1', `2' and `3' increases at the expense of reducing the volume of bricks with higher labels.
This means a subsequent reduction of the number of opinions available in the system.
Unfortunately (for computational sociologists), the rate of this reduction is very slow: The snapshots of the phase diagram presented in \Cref{fig:phase_diagram-t1e3,fig:phase_diagram-t1e6} [and also in \Cref{fig:max_no-t1e3,fig:max_no-t1e6}] are separated by three orders of magnitude in the simulation time $t_\textrm{max}$.

In \Cref{fig:f} we see details of the phase diagram presented in \Cref{fig:phase_diagram-t1e5} in terms of the frequency $f$ (in per mille) of ultimately surviving $n_\textrm{o}^\textrm{u}$ opinions after completing $t_{\textrm{max}}=10^5$ MCS for $n_\textrm{o}^\textrm{u}=1$ [\Cref{fig:f_no_1}], $n_\textrm{o}^\textrm{u}=2$ [\Cref{fig:f_no_2}], $n_\textrm{o}^\textrm{u}=3$ [\Cref{fig:f_no_3}], $n_\textrm{o}^\textrm{u}=4$ [\Cref{fig:f_no_4}], $n_\textrm{o}^\textrm{u}=5$ [\Cref{fig:f_no_5}] and more than five opinions ($n_\textrm{o}^\textrm{u}\ge 6$) [\Cref{fig:f_no_6}]. 
As we can see in \Cref{fig:f_no_1} the social temperature $T\approx 1$ is conducive to reaching consensus as we observe $f>0$ even for $\alpha>4$.
On the other hand, the comparison of \Cref{fig:f_no_1}, \Cref{fig:f_no_2} and \Cref{fig:f_no_3} shows that for $0.95\le T\le 1$ and $5\le\alpha\le 6$ the chance of system polarization outperforms the chance of reaching consensus although surviving of three opinions in this region is the most probable.

%% ============================================================
\begin{figure}[htbp]
%% ------------------------------------------------------------
\begin{subfigure}[b]{0.48\textwidth}
\caption{\label{fig:T0tau}}
\includegraphics[width=.99\textwidth]{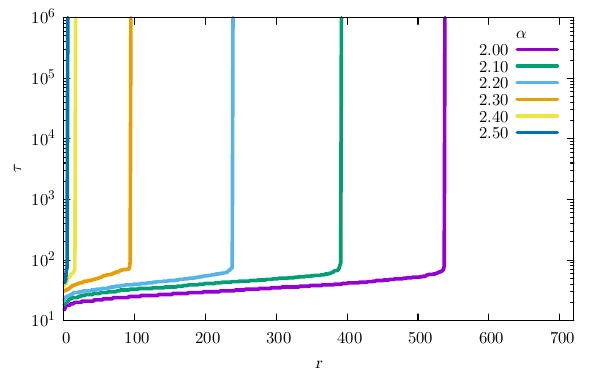}
\end{subfigure}
%% ------------------------------------------------------------
\begin{subfigure}[b]{0.48\textwidth}
\caption{\label{fig:T0hist}}
\includegraphics[width=.99\textwidth]{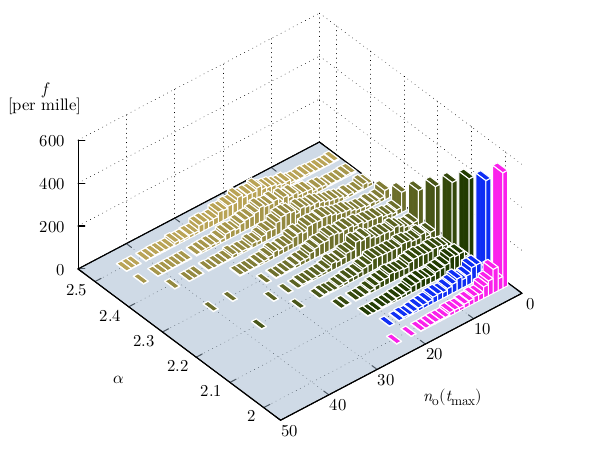}
\end{subfigure}
%% ------------------------------------------------------------
\caption{\label{fig:T0}
Dependencies of \subref{fig:T0tau} the time $\tau$ of reaching the consensus and \subref{fig:T0hist}~distribution of number of surviving opinions $n^\textrm{u}_\textrm{o}$ for deterministic case ($T=0$).
The simulations are carried out until $t_\textrm{max}=10^6$ MCS are performed and the results are averaged over $R=10^3$ simulations.
\subref{fig:T0tau}~The time $\tau$ of reaching the consensus as dependent on the number $r$ of the performed simulation (ranked in ascending order) for various values of $\alpha$. The fraction of simulations leading to $\tau\le t_\textrm{max}=10^6$ monotonically decreases with the effective range of interactions expressed by the values of $\alpha$. 
\subref{fig:T0hist}~Distribution of final number of opinions $n^\textrm{u}_\textrm{o}$ as dependent on the effective range of interactions $\alpha$. With increase of $\alpha$ probability of reaching the consensus decreases}
\end{figure}
%% ============================================================

\Cref{fig:tau} shows time $\tau$ of reaching the consensus as dependent on the number of simulations (here with numeric label $r$ of the simulation sorted accordingly to the increasing time $\tau$ of reaching the consensus) for $\alpha=2$ [\Cref{fig:tau_alpha2}], $\alpha=3$ [\Cref{fig:tau_alpha3}], $\alpha=4$ [\Cref{fig:tau_alpha4}] and various temperatures $T$.
As we can see, the times $\tau$ of reaching consensus are limited by the assumed maximal simulation time (here $t_\textrm{max}=10^6$) for $\alpha=2$ and $T<1$ [\Cref{fig:tau_alpha2}], $\alpha=3$ except for $1.5<T<1.7$ [\Cref{fig:tau_alpha3}] and for all temperatures $T$ presented for $\alpha=4$ [\Cref{fig:tau_alpha4}]. 
The similar restriction of time to reach the consensus $\tau\le t_\textrm{max}=10^6$ is also observed for the deterministic version of the algorithm (for $T=0$) as presented in \Cref{fig:T0tau} for various values of $\alpha$.
The fraction of simulations leading to $\tau\le t_\textrm{max}=10^6$ monotonically decreases with the effective range of interactions expressed by the values of $\alpha$. 
This is even more apparent in \Cref{fig:T0hist}, where the distribution of $n^\textrm{u}_\textrm{o}$ is presented.
The increase in the parameter $\alpha$ reduces the effective range of interaction, which diminishes the chance of reaching a consensus.

In the non-deterministic case ($T>0$), for a finite system (finite $L$) the presence of Muller's ratchet in the model rules [restriction \eqref{eq:prob_Ieq0}] makes the probability of any opinion vanishing finite.
In principle, it is only a matter of time that just one opinion survives.
However, the time to reach the consensus in Latan\'e model seems to be extremely long.
When Muller's ratchet is excluded from the model rules [absence of restriction \eqref{eq:prob_Ieq0}], at high temperatures ($T\to\infty$) the appearance of every opinion $\Lambda_k$ becomes equally probable, and its abundance in the system in the limit of $t\to\infty$ is $L^2/K$ \cite{1902.03454}.

In the deterministic version of the algorithm ($T=0$) the situation is quite opposite: the stable (long-lived states) of the system with $n^\textrm{u}_\textrm{o}>1$ are possible as shown by \citeauthor{Lewenstein_1992} in Reference~\onlinecite{Lewenstein_1992}.
Examples of such states are presented in \Cref{fig:frozen} in \Cref{app:long-times}.

Therefore, at the lowest temperatures ($T\to 0$) we observe remnants of this stability and a multitude of observed opinions.
However, even after $10^6$ MCS the non-zero probability of changes in the state of the system is observed.
In \Cref{fig:Psus} in \Cref{app:long-times} we show examples of maps of opinions for $t_\textrm{max}=10^6$ (in the left column) and associated probabilities $P$ of sustaining opinions (in the right column).
As one may expect, these probabilities are finite ($P<1$) at the boundaries between various opinions.

It seems that the most intriguing result is the insensibility of the largest number of surviving opinions ($\approx 55$, 34 and 29 for $t_{\text{max}}=10^3$, $10^5$ and $10^6$, respectively) on the parameters $\alpha$ and $T$ when they are high enough (see upper right corners in \Cref{fig:max_no}). 
The border line of appearance of these numbers on the maps presented in \Cref{fig:max_no} is also clearly visible on the maps of frequencies $f$ of ultimately surviving opinions (\Cref{fig:f}) for $n^\text{u}_\text{o}=1$ [\Cref{fig:f_no_1}] and $n^\text{u}_\text{o}>5$ [\Cref{fig:f_no_6}] but totally undetectable for maps for $2\le n^\text{u}_\text{o}\le 5$ [\Cref{fig:f_no_2,fig:f_no_3,fig:f_no_4,fig:f_no_5}].

%% ############################################################
\section{\label{sec:conclusion}Conclusion}
%% ############################################################

In this paper, the opinion dynamics model based on the social impact theory of Latan\'e enriched with Muller's ratchet is reconsidered.
With computer simulation, we check the time evolution of the phase diagram for this model, when the fully differentiated society at initial time is assumed (that is, every actor starts with their own opinion).

When the observation time $t_{\text{max}}$ increases, consensus is reached in a systematically wider range of parameters $(\alpha,T)$. 
However, this consensus is only partial in some cases, depending on the exact position in the $(\alpha,T)$-space.
Except for the lowest studied values of the parameter $\alpha$ the characteristic pattern of the thermal evolution is observed: for both low and high temperature the phase labeled `6' prevails.
However, the sources of this prevalence have totally different grounds.
For low values of $T$ the system is `frozen' far from consensus, while for high temperatures the Boltzmann-like factors \eqref{eq:prob_Igt0} for selecting any of still available opinions become roughly equal, although the number of available opinions decreases.

It is clear that the possibility of reaching consensus is limited only by the assumed simulation time $t_{\text{max}}$ (in our case set to $10^3$, $10^5$ and $10^6$ MCS).
The further extension of this time, let us say for next decade, that is up to $t_{\text{max}}=10^7$---even for such moderate system size as $L^2=441$ actors---excludes possibility of accomplishing simulations in a reasonable real-world time, even with parallelization of code and access to TOP500 most powerful supercomputers.
In our opinion, the system governed by the theory of social impact in the presence of finite social temperature $T>0$ ultimately tends to consensus.
However, the time to reach this consensus is extremely long even for relatively small system sizes.

In contrast to earlier approaches \cite{1902.03454,2002.05451,2211.04183}, in this study we maintain the genetically motivated sociological equivalent of Muller's ratchet \cite{Muller_1932,Tiggemann1998} introduced in Reference~\onlinecite{Maslyk_2023}.
As we deal with finite-size systems, the probability of vanishing of any opinion $\Lambda_k$ ($k=1,\cdots,K$) available in the system is also finite.
In other words, it is only a matter of time when all---except one---opinions will disappear, and ultimately the consensus will take place.
In contrast, for the deterministic version stable clusters of various opinions emerge.

After $t_\textrm{max}=10^5$ MCS for $\alpha=6$ and $1\le T\le 1.05$, and also for $\alpha=5$ and $T=1$, we observe $f(n_\textrm{o}^\textrm{u}=2)>f(n_\textrm{o}^\textrm{u}=1)$, which means that in this range of parameters the system polarization is more probable than reaching consensus. 
We conclude that the intermediate social noise $T\approx 1$ and low effective range of interaction $\alpha>4$ favor opinion polarization in society.

%% ############################################################
\begin{acknowledgments}
We gratefully acknowledge Polish high-performance computing infrastructure \href{https://ror.org/01m3qaz74}{PLGrid} (HPC Center: ACK Cyfronet AGH) for providing computer facilities and support within computational grant no.~PLG/2024/017607.
\end{acknowledgments}
%% ############################################################

%% ############################################################
%% \bibliography{km,opiniondynamics,this}
%% ############################################################


%% \clearpage
\appendix
%% ############################################################
%% ############################################################

%% ############################################################
\section{\label{app:small}Examples of small system evolution}
%% ############################################################

%% ============================================================
\begin{figure}[htbp]
%% ------------------------------------------------------------
\begin{subfigure}[b]{0.235\textwidth}
\caption{\label{fig:example-pi}}
\includegraphics[width=.99\textwidth]{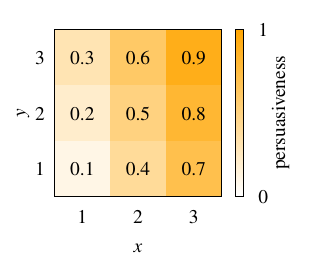}
\end{subfigure}
%% ------------------------------------------------------------
\hfill
%% ------------------------------------------------------------
\begin{subfigure}[b]{0.235\textwidth}
\caption{\label{fig:example-si}}
\includegraphics[width=.99\textwidth]{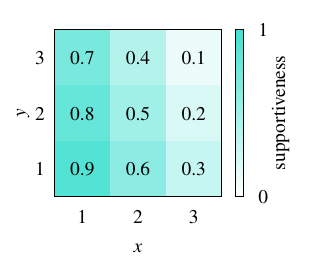}
\end{subfigure}
%% ------------------------------------------------------------
\caption{\label{fig:example-si-pi}Example of actors' \subref{fig:example-pi} persuasiveness $p_{x,y}$ and \subref{fig:example-si} supportiveness $s_{x,y}$ for small system of nine actors}
\end{figure}
%% ============================================================

\begin{figure*}[htbp]
%% ------------------------------------------------------------
\begin{subfigure}[b]{0.20\textwidth}
\caption{\label{fig:opinions_t0}$\lambda_{(x,y)}$, $t=0$}
\includegraphics[width=.99\textwidth]{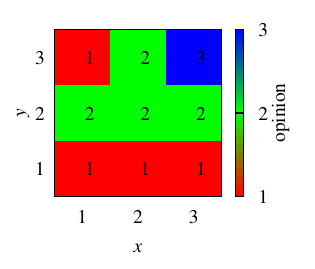}    
\end{subfigure}
\hfill
\begin{subfigure}[b]{0.20\textwidth}
\caption{\label{fig:impact_R_t0}$\mathcal I_{(x,y),R}$, $t=0$}
\includegraphics[width=.99\textwidth]{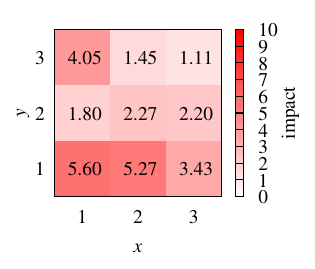}
\end{subfigure}
\hfill
\begin{subfigure}[b]{0.20\textwidth}
\caption{\label{fig:impact_G_t0}$\mathcal I_{(x,y),G}$, $t=0$}
\includegraphics[width=.99\textwidth]{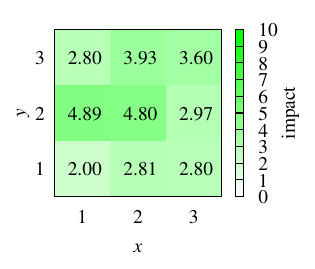}
\end{subfigure}
\hfill
\begin{subfigure}[b]{0.20\textwidth}
\caption{\label{fig:impact_B_t0}$\mathcal I_{(x,y),B}$, $t=0$}
\includegraphics[width=.99\textwidth]{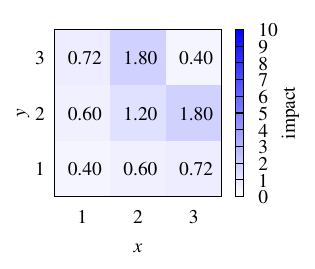}
\end{subfigure}
%% ------------------------------------------------------------
\begin{subfigure}[b]{0.20\textwidth}
\caption{\label{fig:opinions_t1}$\lambda_{(x,y)}$, $t=1$}
\includegraphics[width=.99\textwidth]{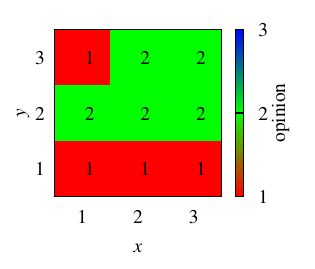}
\end{subfigure}
\hfill
\begin{subfigure}[b]{0.20\textwidth}
\caption{\label{fig:impact_R_t1}$\mathcal I_{(x,y),R}$, $t=1$}
\includegraphics[width=.99\textwidth]{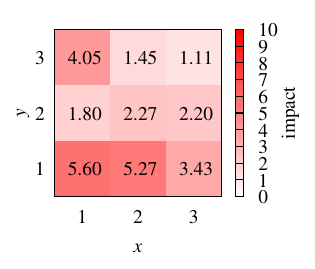}
\end{subfigure}
\hfill
\begin{subfigure}[b]{0.20\textwidth}
\caption{\label{fig:impact_G_t1}$\mathcal I_{(x,y),G}$, $t=1$}
\includegraphics[width=.99\textwidth]{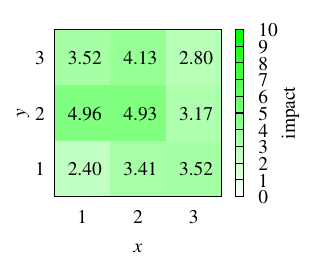}
\end{subfigure}
\hfill
\begin{subfigure}[b]{0.20\textwidth}
\caption{\label{fig:impact_B_t1}$\mathcal I_{(x,y),B}$, $t=1$}
\includegraphics[width=.99\textwidth]{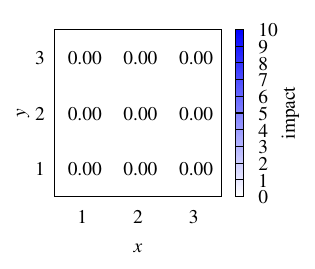}
\end{subfigure}
%% ------------------------------------------------------------
\begin{subfigure}[b]{0.20\textwidth}
\caption{\label{fig:opinions_t2}$\lambda_{(x,y)}$, $t=2$}
\includegraphics[width=.99\textwidth]{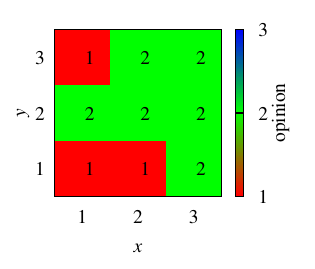}
\end{subfigure}
\hfill
\begin{subfigure}[b]{0.20\textwidth}
\caption{\label{fig:impact_R_t2}$\mathcal I_{(x,y),R}$, $t=2$}
\includegraphics[width=.99\textwidth]{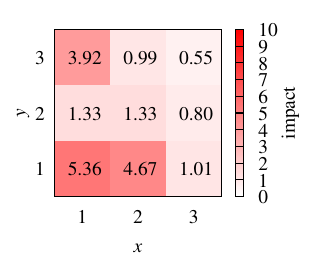}
\end{subfigure}
\hfill
\begin{subfigure}[b]{0.20\textwidth}
\caption{\label{fig:impact_G_t2}$\mathcal I_{(x,y),G}$, $t=2$}
\includegraphics[width=.99\textwidth]{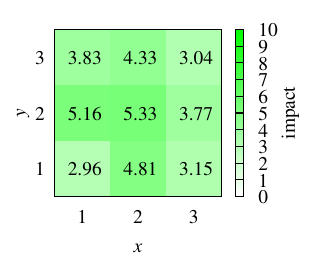}
\end{subfigure}
\hfill
\begin{subfigure}[b]{0.20\textwidth}
\caption{\label{fig:impact_B_t2}$\mathcal I_{(x,y),B}$, $t=2$}
\includegraphics[width=.99\textwidth]{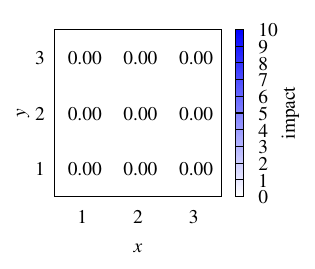}
\end{subfigure}
%% ------------------------------------------------------------
\begin{subfigure}[b]{0.20\textwidth}
\caption{\label{fig:opinions_t3}$\lambda_{(x,y)}$, $t=3$}
\includegraphics[width=.99\textwidth]{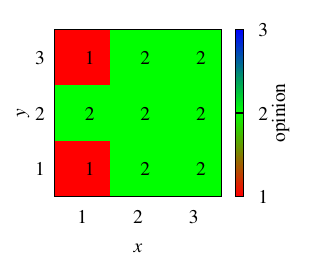}
\end{subfigure}
\hfill
\begin{subfigure}[b]{0.20\textwidth}
\caption{\label{fig:impact_R_t3}$\mathcal I_{(x,y),R}$, $t=3$}
\includegraphics[width=.99\textwidth]{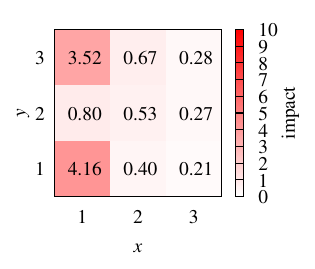}
\end{subfigure}
\hfill
\begin{subfigure}[b]{0.20\textwidth}
\caption{\label{fig:impact_G_t3}$\mathcal I_{(x,y),G}$, $t=3$}
\includegraphics[width=.99\textwidth]{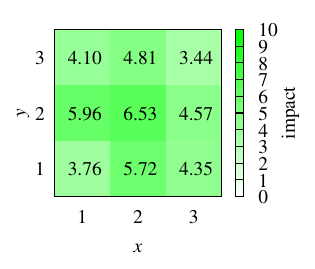}
\end{subfigure}
\hfill
\begin{subfigure}[b]{0.20\textwidth}
\caption{\label{fig:impact_B_t3}$\mathcal I_{(x,y),B}$, $t=3$}
\includegraphics[width=.99\textwidth]{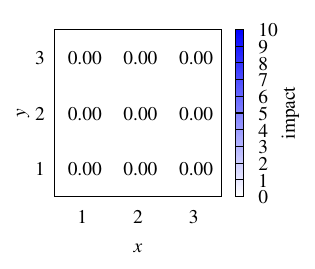}
\end{subfigure}
%% ------------------------------------------------------------
\begin{subfigure}[b]{0.20\textwidth}
\caption{\label{fig:opinions_t4}$\lambda_{(x,y)}$, $t=4$}
\includegraphics[width=.99\textwidth]{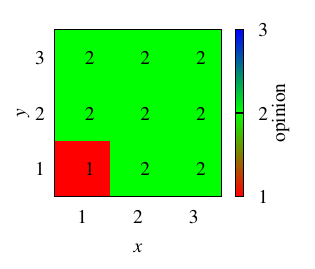}
\end{subfigure}
\hfill
\begin{subfigure}[b]{0.20\textwidth}
\caption{\label{fig:impact_R_t4}$\mathcal I_{(x,y),R}$, $t=4$}
\includegraphics[width=.99\textwidth]{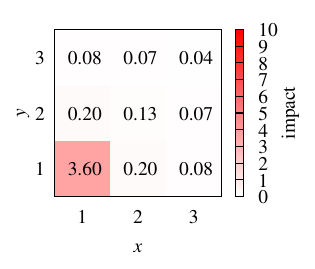}
\end{subfigure}
\hfill
\begin{subfigure}[b]{0.20\textwidth}
\caption{\label{fig:impact_G_t4}$\mathcal I_{(x,y),G}$, $t=4$}
\includegraphics[width=.99\textwidth]{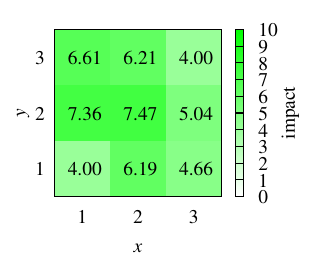}
\end{subfigure}
\hfill
\begin{subfigure}[b]{0.20\textwidth}
\caption{\label{fig:impact_B_t4}$\mathcal I_{(x,y),B}$, $t=4$}
\includegraphics[width=.99\textwidth]{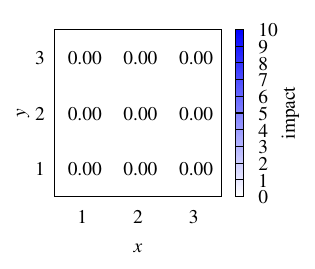}
\end{subfigure}
%% -----------------------------------------------------------
\begin{subfigure}[b]{0.20\textwidth}
\caption{\label{fig:opinions_t5}$\lambda_{(x,y)}$, $t=5$}
\includegraphics[width=.99\textwidth]{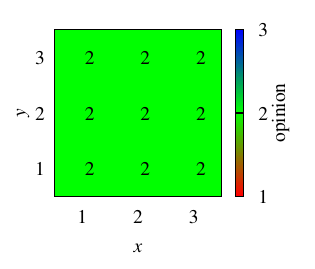}
\end{subfigure}
\hfill
\begin{subfigure}[b]{0.20\textwidth}
\caption{\label{fig:impact_R_t5}$\mathcal I_{(x,y),R}$, $t=5$}
\includegraphics[width=.99\textwidth]{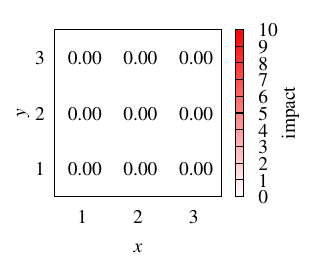}
\end{subfigure}
\hfill
\begin{subfigure}[b]{0.20\textwidth}
\caption{\label{fig:impact_G_t5}$\mathcal I_{(x,y),G}$, $t=5$}
\includegraphics[width=.99\textwidth]{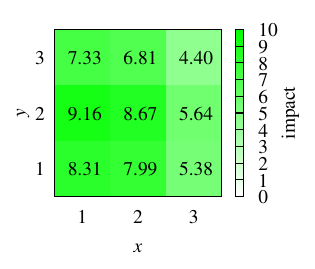}
\end{subfigure}
\hfill
\begin{subfigure}[b]{0.20\textwidth}
\caption{\label{fig:impact_B_t5}$\mathcal I_{(x,y),B}$, $t=5$}
\includegraphics[width=.99\textwidth]{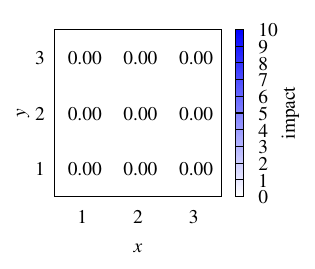}
\end{subfigure}
%% ------------------------------------------------------------
\caption{\label{fig:example}Example of time evolution of opinions in small system of nine actors. Their opinions $\lambda_{xy}$ are presented in the first column. The second, third, and fourth columns indicate social impacts $\mathcal I_{(x,y),C}$ for opinion $C$ equal to `red', `green' and `blue' opinions, respectively}
\end{figure*}

Let us calculate some impacts $\mathcal{I}$ for a toy system of nine actors with $K=3$ opinions marked `red' ($R$), `green' ($G$) and `blue' ($B$) and perform step-by-step system evolution with the deterministic version of the algorithm ($T=0$).  
Actors' supportiveness $s_{x,y}$ and persuasiveness $p_{x,y}$ are indicated on \Cref{fig:example-si-pi}.
We assume $\alpha=2$ in the distance scaling function \eqref{eq:fg}.
Here, $\mathcal I_{(x,y);C}$ represents the social impact on the actor in the position $(x,y)$ exerted by the actors who have the opinion $C$.
Let us start with calculation of the social impact exerted by believers of each opinion available in the system at three arbitrarily selected positions $(1,1)$, $(1,3)$ and $(3,3)$.
Initially, at $t=0$, the actors have opinions presented in \Cref{fig:opinions_t0}.

According to \Cref{eq:szamrej_sum_same}, the impact of (the single) believer of opinion $B$ at position $(3,3)$ and at time $t=0$ is
%% ------------------------------------------------------------
\begin{equation} 
\mathcal{I}_{(3,3);B}=4 \left(\frac{s_{3,3}}{1+d_{(3,3);(3,3)}^2}\right) = \frac{4 \cdot 0.1}{1+0^2}  = 0.4
\end{equation}
%% ------------------------------------------------------------
(since any single believer has no more supporters than themself) and at coordinates $(1,1)$, $(1,3)$---according to \Cref{eq:szamrej_sum_diff}
%% ------------------------------------------------------------
\begin{equation}
\mathcal{I}_{(1,1);B}=4 \left(\frac{p_{3,3}}{1+d_{(3,3);(1,1)}^2}\right) = \frac{4 \cdot 0.9}{1+(2\sqrt{2})^2} = 0.4,
\end{equation}
%% ------------------------------------------------------------
and
%% ------------------------------------------------------------
\begin{equation}
\mathcal{I}_{(1,3);B}=4 \left(\frac{p_{3,3}}{1+d_{(3,3);(1,3)}^2}\right) = \frac{4 \cdot 0.9}{1+2^2} = 0.72.
\end{equation}
%% ------------------------------------------------------------

\begin{widetext}
The impacts on these three positions by other opinions (`red' and `green') at time $t=0$ are
%% ------------------------------------------------------------
\begin{equation}
\begin{split}
\mathcal{I}_{(3,3);R}
=4\left(\frac{p_{1,3}}{1+d_{(3,3);(1,3)}^2} + \frac{p_{1,1}}{1+d_{(3,3);(1,1)}^2} + \frac{p_{2,1}}{1+d_{(3,3);(2,1)}^2} + \frac{p_{3,1}}{1+d_{(3,3);(3,1)}^2} \right) \\
=4\left(\frac{0.3}{1+2^2} + \frac{0.1}{1+(2\sqrt{2})^2} + \frac{0.4}{1+\sqrt{5}^2} + \frac{0.7}{1+2^2}\right)\approx 1.11, 
\end{split}
\end{equation}
%% ------------------------------------------------------------
\begin{equation}
\begin{split}
\mathcal{I}_{(3,3);G}
=4\left(\frac{p_{2,3}}{1+d_{(3,3);(2,3)}^2} + \frac{p_{1,2}}{1+d_{(3,3);(1,2)}^2} + \frac{p_{2,2}}{1+d_{(3,3);(2,2)}^2} + \frac{p_{3,2}}{1+d_{(3,3);(3,2)}^2}\right) \\
=4\left(\frac{0.6}{1+1^2} + \frac{0.2}{1+\sqrt{5}^2} + \frac{0.5}{1+\sqrt{2}^2} + \frac{0.8}{1+1^2}\right) = 3.6,  
\end{split}
\end{equation}
%% ------------------------------------------------------------
\begin{equation}
\begin{split}
\mathcal{I}_{(1,1);R}=4 \left(\frac{s_{1,1}}{1+d_{(1,1);(1,1)}^2} + 
\frac{s_{1,3}}{1+d_{(1,1);(1,3)}^2} + 
\frac{s_{2,1}}{1+d_{(1,1);(2,1)}^2} + 
\frac{s_{3,1}}{1+d_{(1,1);(3,1)}^2}\right)\\
 =4 \left(\frac{0.9}{1+0^2} + 
\frac{0.7}{1+2^2} + 
\frac{0.6}{1+1^2} + 
\frac{0.3}{1+2^2}\right)
= 5.6, 
\end{split}
\end{equation}
%% ------------------------------------------------------------
\begin{equation}
\begin{split}
\mathcal{I}_{(1,1);G}=4 \left(\frac{p_{1,2}}{1+d_{(1,1);(1,2)}^2} + 
\frac{p_{2,2}}{1+d_{(1,1);(2,2)}^2} + 
\frac{p_{2,3}}{1+d_{(1,1);(2,3)}^2} + 
\frac{p_{3,2}}{1+d_{(1,1);(3,2)}^2}\right)\\
= 4 \left(\frac{0.2}{1+1^2} + 
\frac{0.5}{1+\sqrt{2}^2} + 
\frac{0.6}{1+\sqrt{5}^2} + 
\frac{0.8}{1+\sqrt{5}^2}\right)
= 2,
\end{split}
\end{equation}
%% ------------------------------------------------------------
\begin{equation}
\begin{split} 
\mathcal{I}_{(1,3);R}=4 \left(\frac{s_{1,3}}{1+d_{(1,3);(1,3)}^2} + 
\frac{s_{1,1}}{1+d_{(1,3);(1,1)}^2} + 
\frac{s_{2,1}}{1+d_{(1,3);(2,1)}^2} + 
\frac{s_{3,1}}{1+d_{(1,3);(3,1)}^2}\right)\\
 = 4 \left(\frac{0.7}{1+0^2} + 
\frac{0.9}{1+2^2} + 
\frac{0.6}{1+\sqrt{5}^2} + 
\frac{0.3}{1+\sqrt{8}^2}\right)
\approx 4.05,
\end{split}
\end{equation}
%% ------------------------------------------------------------
\begin{equation}
\begin{split}
\mathcal{I}_{(1,3);G}=4 \left(\frac{p_{1,2}}{1+d_{(1,3);(1,2)}^2} + 
\frac{p_{2,2}}{1+d_{(1,3);(2,2)}^2} + 
\frac{p_{3,2}}{1+d_{(1,3);(3,2)}^2} + 
\frac{p_{2,3}}{1+d_{(1,3);(2,3)}^2}\right)\\
 = 4\left(\frac{0.2}{1+1^2} + 
\frac{0.5}{1+\sqrt{2}^2} + 
\frac{0.8}{1+\sqrt{5}^2} + 
\frac{0.6}{1+1^2}\right) 
= 2.8.
\end{split}
\end{equation}
%% ------------------------------------------------------------

Thus, in the next step the actors at positions $(1,1)$ and $(1,3)$ will sustain their `red' opinion as 
$\mathcal{I}_{(1,1);R}>\mathcal{I}_{(1,1);G}>\mathcal{I}_{(1,1);B}$
and
$\mathcal{I}_{(1,3);R}>\mathcal{I}_{(1,3);G}>\mathcal{I}_{(1,3);B}$.
In contrast, the actor at position $(3,3)$ will change their opinion from `blue' to `green'---as $\mathcal{I}_{(3,3);G}>\mathcal{I}_{(3,3);R}>\mathcal{I}_{(3,3);B}$.

The subsequent time steps (up to $t=5$) are presented in the following rows of \Cref{fig:example}.
The first column shows the time evolution of the opinions $\lambda_{(x,y)}$ of actors at sites $(x,y)$, while the second, third, and fourth columns indicate social impacts $\mathcal I_{(x,y);C}$ for opinions $C$ (here colored as: `red', `green' and `blue'), respectively. 

At $t=1$ the impacts on `red' actor at position $(3,1)$ are
%% ------------------------------------------------------------
\begin{equation}
\mathcal{I}_{(3,1);B}=0,
\end{equation}
%% ------------------------------------------------------------
%% ------------------------------------------------------------
\begin{equation}
\begin{split}
\mathcal{I}_{(3,1);R}= 4 \left(
\frac{s_{1,1}}{1+d_{(3,1);(1,1)}^2} + 
\frac{s_{2,1}}{1+d_{(3,1);(2,1)}^2} + 
\frac{s_{3,1}}{1+d_{(3,1);(3,1)}^2} + 
\frac{s_{1,3}}{1+d_{(3,1);(1,3)}^2} \right)\\ =
4 \left(
\frac{0.9}{1+2^2} + 
\frac{0.6}{1+1^2} + 
\frac{0.3}{1+0^2} + 
\frac{0.7}{1+(2\sqrt{2})^2} \right) \approx 3.43,
\end{split}
\end{equation}
%% ------------------------------------------------------------
%% ------------------------------------------------------------
\begin{equation}
\begin{split}
\mathcal{I}_{(3,1);G}= 4 \left(
\frac{p_{1,2}}{1+d_{(3,1);(1,2)}^2} + 
\frac{p_{2,2}}{1+d_{(3,1);(2,2)}^2} + 
\frac{p_{3,2}}{1+d_{(3,1);(3,2)}^2} + 
\frac{p_{2,3}}{1+d_{(3,1);(2,3)}^2} +
\frac{p_{3,3}}{1+d_{(3,1);(3,3)}^2} \right)\\ =
4 \left(
\frac{0.2}{1+\sqrt{5}^2} + 
\frac{0.5}{1+\sqrt{2}^2} + 
\frac{0.8}{1+1^2} + 
\frac{0.6}{1+\sqrt{5}^2} +
\frac{0.9}{1+2^2} \right) = 3.52
\end{split}
\end{equation}
%% ------------------------------------------------------------
and as $\mathcal{I}_{(3,1);G} > \mathcal{I}_{(3,1);R} > \mathcal{I}_{(3,1);B}$ the actor at site $(3,1)$ changes their opinion from `red' (\Cref{fig:opinions_t1}) to `green' (\Cref{fig:opinions_t2}).

At $t=2$ the impacts on `red' actor at position $(2,1)$ are
%% ------------------------------------------------------------
\begin{equation}
\mathcal{I}_{(2,1);B}=0,
\end{equation}
%% ------------------------------------------------------------
%% ------------------------------------------------------------
\begin{equation}
\begin{split}
\mathcal{I}_{(2,1);R}= 4 \left(
\frac{s_{1,1}}{1+d_{(2,1);(1,1)}^2} + 
\frac{s_{2,1}}{1+d_{(2,1);(2,1)}^2} + 
\frac{s_{1,3}}{1+d_{(2,1);(1,3)}^2} \right)\\ =
4 \left(
\frac{0.9}{1+1^2} + 
\frac{0.6}{1+0^2} + 
\frac{0.7}{1+\sqrt{5}^2} \right) \approx 4.67,
\end{split}
\end{equation}
%% ------------------------------------------------------------
%% ------------------------------------------------------------
\begin{equation}
\begin{split}
\mathcal{I}_{(2,1);G}= 4 \left(
\frac{p_{3,1}}{1+d_{(2,1);(3,1)}^2} + 
\frac{p_{1,2}}{1+d_{(2,1);(1,2)}^2} + 
\frac{p_{2,2}}{1+d_{(2,1);(2,2)}^2} + 
\frac{p_{3,2}}{1+d_{(2,1);(3,2)}^2} + 
\frac{p_{2,3}}{1+d_{(2,1);(2,3)}^2} +
\frac{p_{3,3}}{1+d_{(2,1);(3,3)}^2} \right)\\ =
4 \left(
\frac{0.7}{1+1^2} + 
\frac{0.2}{1+\sqrt{2}^2} + 
\frac{0.5}{1+1^2} + 
\frac{0.8}{1+\sqrt{2}^2} + 
\frac{0.6}{1+2^2} +
\frac{0.9}{1+\sqrt{5}^2} \right) \approx 4.81
\end{split}
\end{equation}
%% ------------------------------------------------------------
and as $\mathcal{I}_{(2,1),G} > \mathcal{I}_{(2,1),R} > \mathcal{I}_{(2,1),B}$ the actor at site $(2,1)$ changes their opinion from `red' (\Cref{fig:opinions_t2}) to `green' (\Cref{fig:opinions_t3}).

At $t=3$ the impacts on `red' actor at position $(1,3)$ are
%% ------------------------------------------------------------
\begin{equation}
\mathcal{I}_{(1,3);B}=0,
\end{equation}
%% ------------------------------------------------------------
%% ------------------------------------------------------------
\begin{equation}
\begin{split}
\mathcal{I}_{(1,3);R}=4 \left(\frac{s_{1,3}}{1+d_{(1,3);(1,3)}^2} + \frac{s_{1,1}}{1+d_{(1,1);(1,3)}^2}\right)
=4\left(\frac{0.7}{1+0^2} + \frac{0.9}{1+2^2}\right) = 3.52,
\end{split}
\end{equation}
%% ------------------------------------------------------------
%% ------------------------------------------------------------
\begin{equation}
\begin{split}
\mathcal{I}_{(1,3);G}=4\Bigg(
\frac{p_{2,1}}{1+d_{(1,3);(2,1)}^2} + 
\frac{p_{3,1}}{1+d_{(1,3);(2,3)}^2} + 
\frac{p_{1,2}}{1+d_{(1,3);(1,2)}^2} + 
\frac{p_{2,2}}{1+d_{(1,3);(2,2)}^2} +\\
\frac{p_{3,2}}{1+d_{(1,3);(3,2)}^2} + 
\frac{p_{2,3}}{1+d_{(1,3);(2,3)}^2} + 
\frac{p_{3,3}}{1+d_{(1,3);(3,3)}^2} \Bigg) \\
=4\left(\frac{0.4}{1+\sqrt{5}^2} + \frac{0.7}{1+(2\sqrt{2})^2} + \frac{0.2}{1+1^2} + \frac{0.5}{1+\sqrt{2}^2} + \frac{0.8}{1+\sqrt{5}^2} + \frac{0.6}{1+1^2} + \frac{0.9}{1+2^2}\right) \approx 4.10
\end{split}
\end{equation}
%% ------------------------------------------------------------
and as $\mathcal{I}_{(1,3);G} > \mathcal{I}_{(1,3);R} > \mathcal{I}_{(1,3);B}$ the actor at site $(1,3)$ changes their opinion from `red' (\Cref{fig:opinions_t3}) to `green' (\Cref{fig:opinions_t4}).

At $t=4$ the impacts on `red' actor at position $(1,1)$ are
%% ------------------------------------------------------------
\begin{equation}
\mathcal{I}_{(1,1);B}=0,
\end{equation}
%% ------------------------------------------------------------
%% ------------------------------------------------------------
\begin{equation}
\begin{split}
\mathcal{I}_{(1,1);R}=4 \left(\frac{s_{1,1}}{1+d_{(1,1);(1,1)}^2}\right) =
4\left(\frac{0.9}{1+0^2}\right) 
= 3.6,
\end{split}
\end{equation}
%% ------------------------------------------------------------
%% ------------------------------------------------------------
\begin{equation}
\begin{split}
\mathcal{I}_{(1,1);G}=4\Bigg(
\frac{p_{2,1}}{1+d_{(1,1);(2,1)}^2} + 
\frac{p_{2,3}}{1+d_{(1,1);(2,3)}^2} + 
\frac{p_{1,2}}{1+d_{(1,1);(1,2)}^2} + 
\frac{p_{2,2}}{1+d_{(1,1);(2,2)}^2} +\\
\frac{p_{3,2}}{1+d_{(1,1);(3,2)}^2} + 
\frac{p_{1,3}}{1+d_{(1,1);(1,3)}^2} + 
\frac{p_{2,3}}{1+d_{(1,1);(2,3)}^2} + 
\frac{p_{3,3}}{1+d_{(1,1);(3,3)}^2} \Bigg) \\
=4\left(\frac{0.4}{1+1^2} + \frac{0.7}{1+2^2} + \frac{0.2}{1+1^2} + \frac{0.5}{1+\sqrt{2}^2} + \frac{0.8}{1+\sqrt{5}^2} + \frac{0.3}{1+2^2}  + \frac{0.6}{1+\sqrt{5}^2} + \frac{0.9}{1+(2\sqrt{2})^2}\right) = 4
\end{split}
\end{equation}
%% ------------------------------------------------------------
and as $\mathcal{I}_{(1,1);G} > \mathcal{I}_{(1,1);R} > \mathcal{I}_{(1,1);B}$ the actor at site $(1,1)$ changes their opinion from `red' [\Cref{fig:opinions_t4}] to `green' [\Cref{fig:opinions_t5}].
\end{widetext}

%% ============================================================
%% ============================================================

Finally, after completing five time steps, all actors share `green' opinion and the consensus takes place (see \Cref{fig:opinions_t5}).
The presence of a sociological equivalent of Muller's ratchet successfully prevents the restoration of any opinion previously removed from the system.
Thus after eliminating the `blue' opinion, it will never have a chance to appear again, and thus we see zeros in matrices \Cref{fig:impact_B_t1,fig:impact_B_t2,fig:impact_B_t3,fig:impact_B_t4,fig:impact_B_t5} and ultimately also on \Cref{fig:impact_R_t5}---for impact from eliminated `red' opinions.

%% ############################################################
\section{\label{app:distributions}Distribution of numbers of opinions $n_\textrm{o}^\textrm{u}$ observed in the system}
%% ############################################################

\begin{figure*}[htbp]
\includegraphics[width=.73\textwidth]{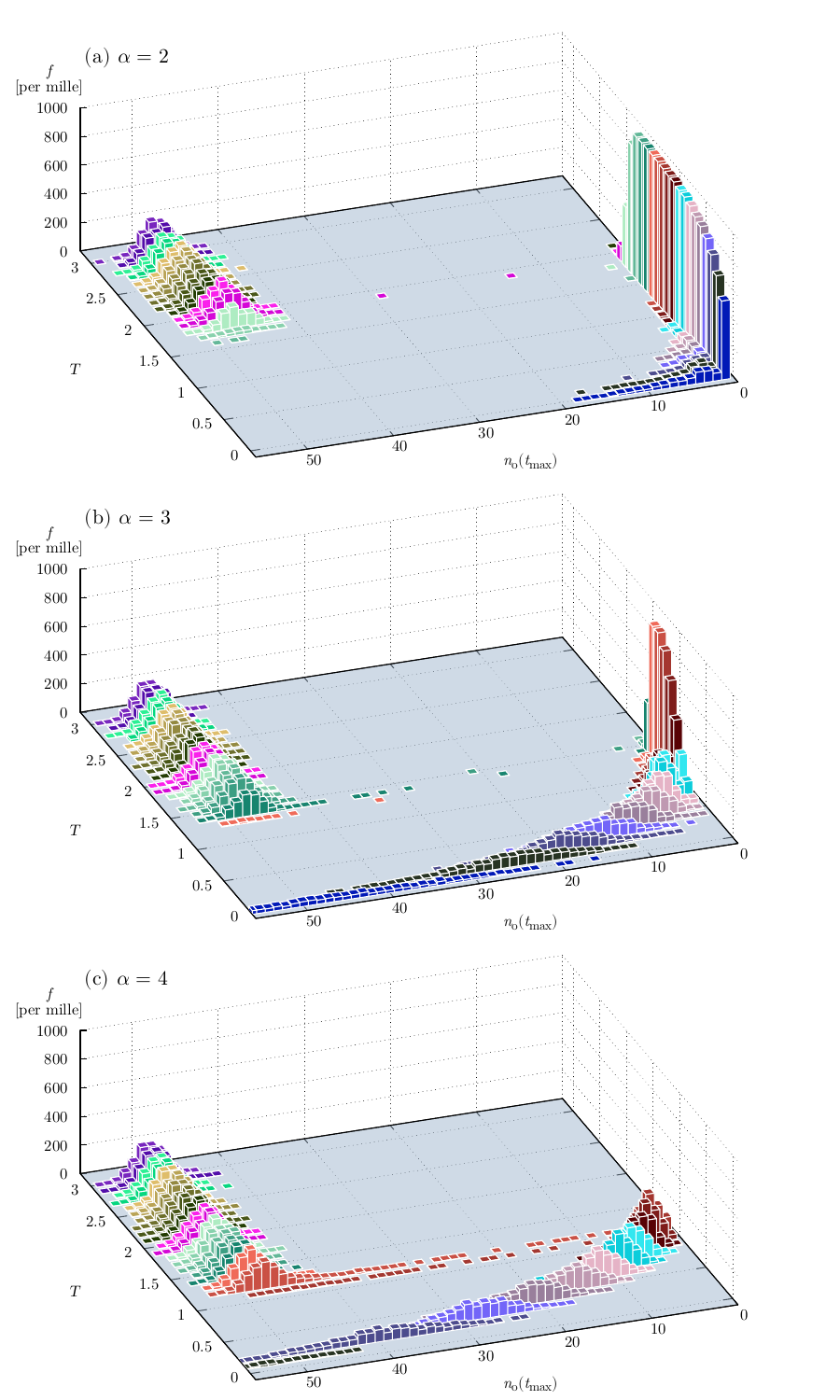}
\caption{\label{fig:no_hist_tmax1e3}Distribution of $n_\textrm{o} (t_\textrm{max}=10^3$). 
(a) $\alpha=2$;
(b) $\alpha=3$, note: $n_\textrm{o}(T=0) \in [16;132]$ (partly visible);
(c) $\alpha=4$, note: $n_\textrm{o}(T=0) \in [104;232]$ (not visible), $n_\textrm{o}(T=0.1) \in [43;115]$ (partly visible)}
\end{figure*}

\begin{figure*}[htbp]
\includegraphics[width=.73\textwidth]{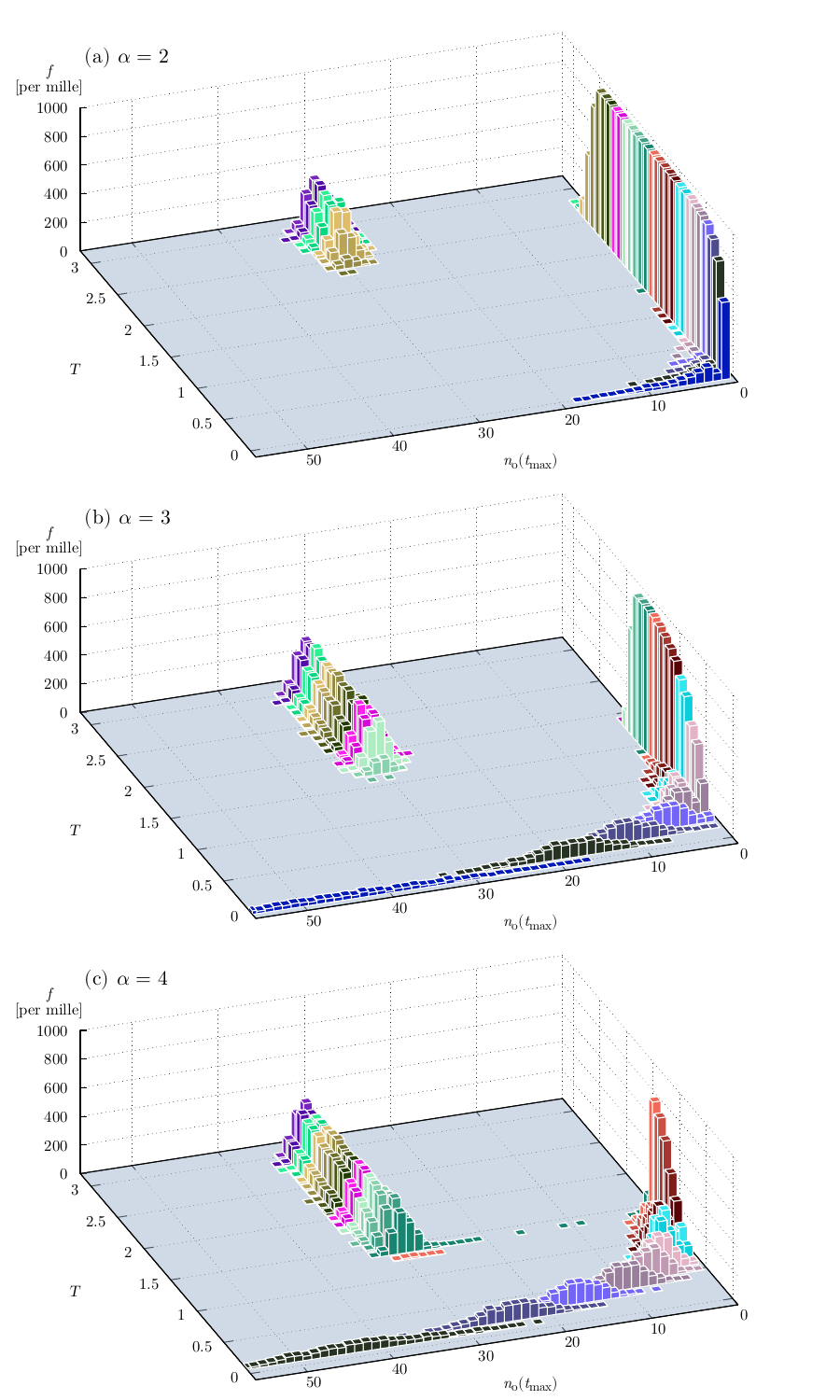}
\caption{\label{fig:no_hist_tmax1e5}Distribution of $n_\textrm{o} (t_\textrm{max}=10^5$). 
(a) $\alpha=2$;
(b) $\alpha=3$, note: $n_\textrm{o}(T=0) \in [17;113]$ (partly visible);
(c) $\alpha=4$, note: $n_\textrm{o}(T=0) \in [100;244]$ (not visible), $n_\textrm{o}(T=0.1) \in [22;70]$ (partly visible)
}
\end{figure*}

\begin{figure*}[htbp]
\includegraphics[width=.73\textwidth]{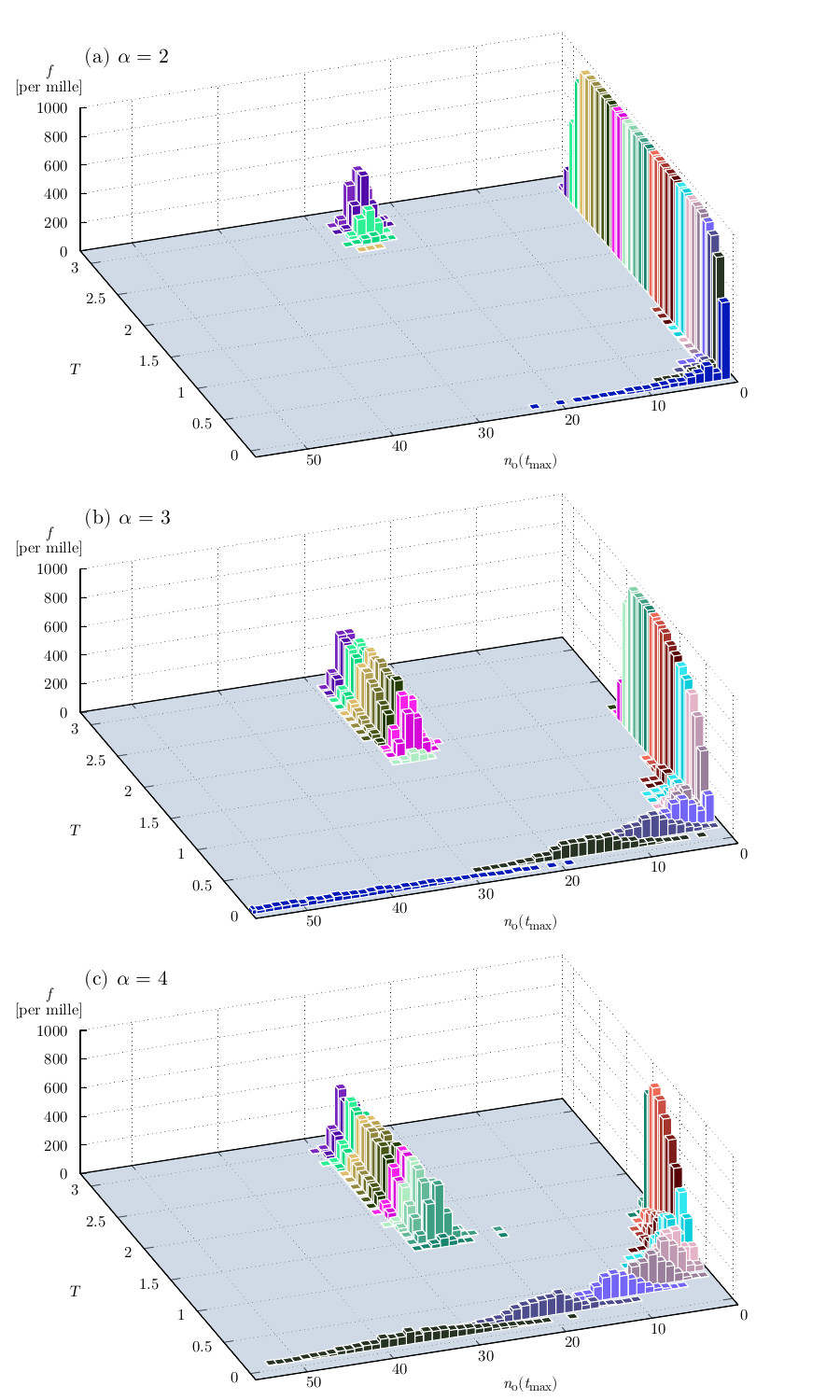}
\caption{\label{fig:no_hist_tmax1e6}Distribution of $n_\textrm{o} (t_\textrm{max}=10^6$). 
(a) $\alpha=2$;
(b) $\alpha=3$, note: $n_\textrm{o}(T=0) \in [19;111]$ (partly visible);
(c) $\alpha=4$, note: $n_\textrm{o}(T=0) \in [105;241]$ (not visible)}
\end{figure*}

\Cref{fig:no_hist_tmax1e3,fig:no_hist_tmax1e5,fig:no_hist_tmax1e6} show detailed distribution of $n_\textrm{o}^\textrm{u}$ on the social temperature $T$ for various parameters $\alpha$ after $t_\textrm{max}=10^3$, $10^5$ and $10^6$, respectively.

%% ============================================================
%% ============================================================

%% ============================================================
%% ============================================================

%% ============================================================
%% ============================================================

%% ############################################################
\section{\label{app:long-times}Examples of long-time system behavior}
\begin{figure*}[htbp]
%% ------------------------------------------------------------
\begin{subfigure}[b]{0.32\textwidth}
\caption{\label{fig:frozen_a2}}
{\includegraphics[trim={6mm 8mm 10mm 8mm},clip,width=\textwidth]{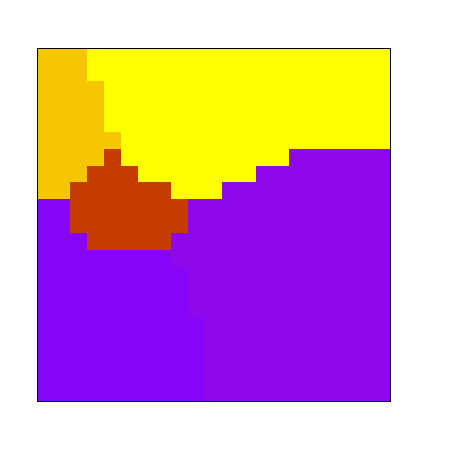}}
\end{subfigure}
\hfill %% ------------------------------------------------------------
\begin{subfigure}[b]{0.32\textwidth}
\caption{\label{fig:frozen_a3}}
{\includegraphics[trim={6mm 8mm 10mm 8mm},clip,width=\textwidth]{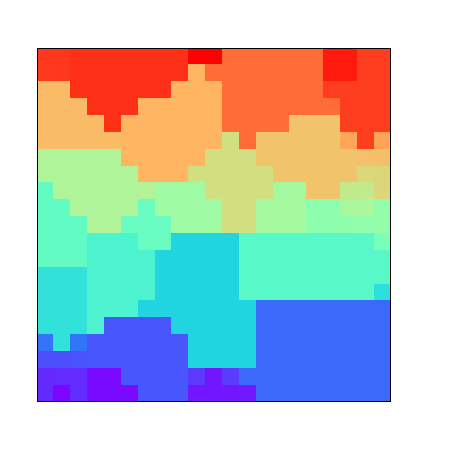}}
\end{subfigure}
\hfill %% ------------------------------------------------------------
\begin{subfigure}[b]{0.32\textwidth}
\caption{\label{fig:frozen_a4}}
{\includegraphics[trim={6mm 8mm 10mm 8mm},clip,width=\textwidth]{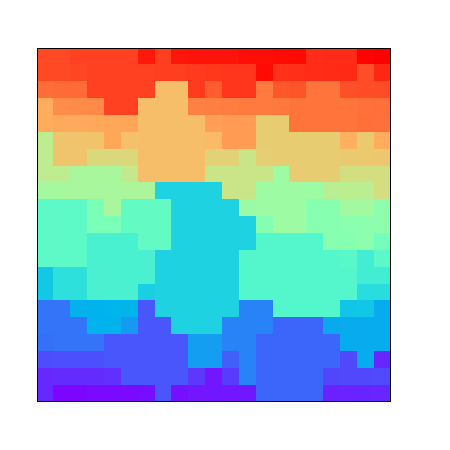}}
\end{subfigure}
%% ------------------------------------------------------------
\caption{\label{fig:frozen}Examples of maps $\lambda$ frozen at $T=0$ for 
\subref{fig:frozen_a2}  $\alpha=2$, $n_\textrm{o}^\textrm{u}=4$,
\subref{fig:frozen_a3} $\alpha=3$, $n_\textrm{o}^\textrm{u}=52$  and 
\subref{fig:frozen_a4} $\alpha=4$, $n_\textrm{o}^\textrm{u}=146$}
\end{figure*}

%% ############################################################
In \Cref{fig:frozen} examples of maps $\lambda$ of opinions frozen in $T=0$ for $\alpha=2$ [\Cref{fig:frozen_a2}], $\alpha=3$ [\Cref{fig:frozen_a3}] and $\alpha=4$ [\Cref{fig:frozen_a4}] are presented.

%% ============================================================
%% ============================================================

\begin{figure*}[htbp]
%% ------------------------------------------------------------
\begin{subfigure}[b]{0.20\textwidth}
\caption{\label{fig:Lambda_a2_T0.5}}
{\includegraphics[trim={3mm 6mm 0mm 6mm},clip,height=30mm]{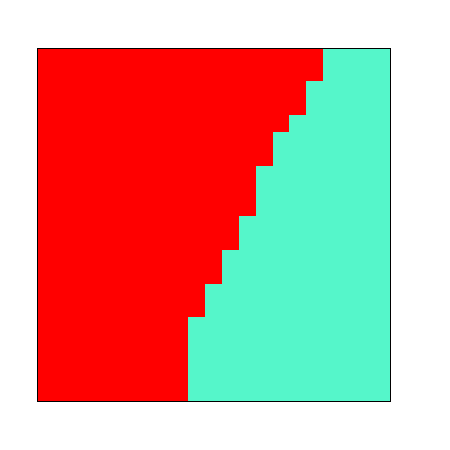}}
\end{subfigure}
%% ------------------------------------------------------------
\begin{subfigure}[b]{0.25\textwidth}
\caption{\label{fig:Prosus_a2_T0.5}}
{\includegraphics[trim={3mm 6mm 0mm 6mm},clip,height=30mm]{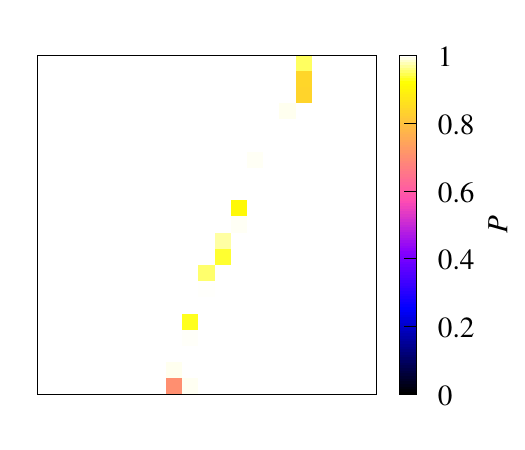}}
\end{subfigure}
\hfill %% ------------------------------------------------------------
\begin{subfigure}[b]{0.20\textwidth}
\caption{\label{fig:Lambda_a2_T1.1}}
{\includegraphics[trim={3mm 6mm 0mm 6mm},clip,height=30mm]{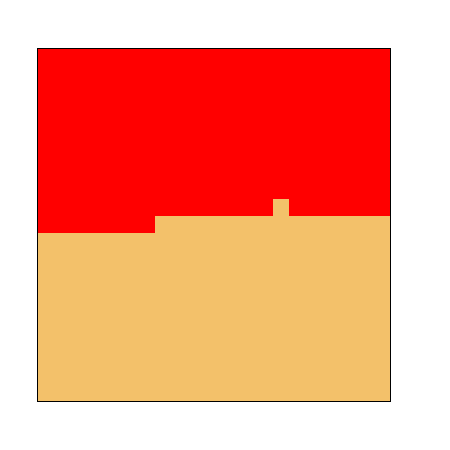}}
\end{subfigure}
%% ------------------------------------------------------------
\begin{subfigure}[b]{0.25\textwidth}
\caption{\label{fig:Prosus_a2_T1.1}}
{\includegraphics[trim={3mm 6mm 0mm 6mm},clip,height=30mm]{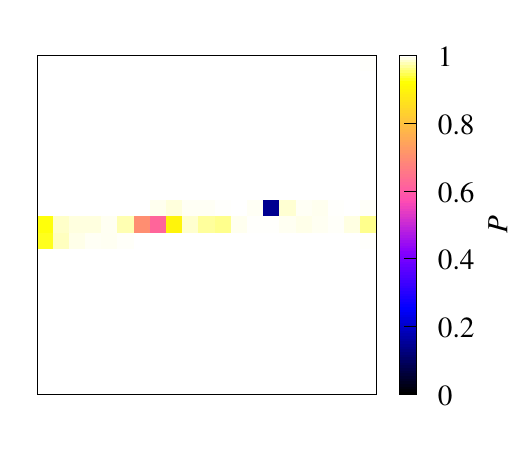}}
\end{subfigure}
%% ------------------------------------------------------------
\begin{subfigure}[b]{0.20\textwidth}
\caption{\label{fig:Lambda_a3_T0.5}}
{\includegraphics[trim={3mm 6mm 0mm 6mm},clip,height=30mm]{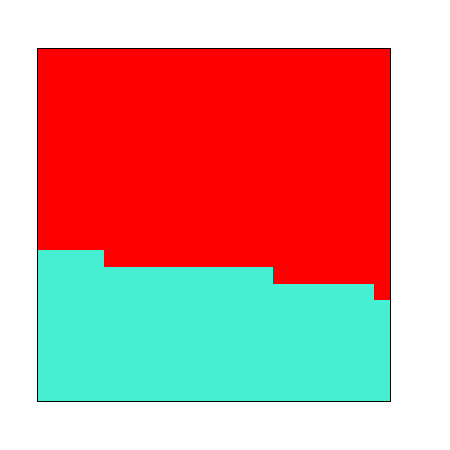}}
\end{subfigure}
%% ------------------------------------------------------------
\begin{subfigure}[b]{0.25\textwidth}
\caption{\label{fig:Prosus_a3_T0.5}}
{\includegraphics[trim={3mm 6mm 0mm 6mm},clip,height=30mm]{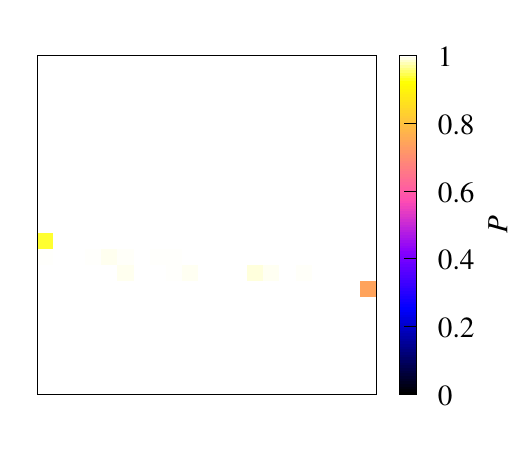}}
\end{subfigure}
\hfill %% ------------------------------------------------------------
\begin{subfigure}[b]{0.20\textwidth}
\caption{\label{fig:Lambda_a3_T1.1}}
{\includegraphics[trim={3mm 6mm 0mm 6mm},clip,height=30mm]{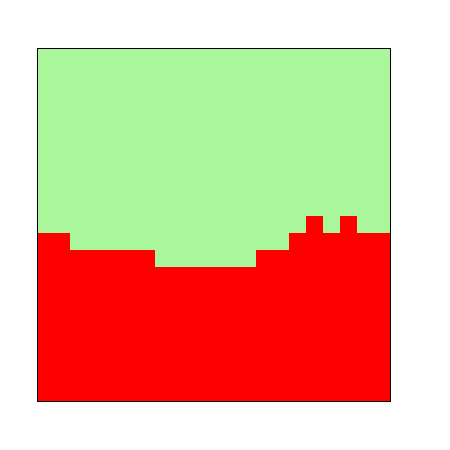}}
\end{subfigure}
%% ------------------------------------------------------------
\begin{subfigure}[b]{0.25\textwidth}
\caption{\label{fig:Prosus_a3_T1.1}}
{\includegraphics[trim={3mm 6mm 0mm 6mm},clip,height=30mm]{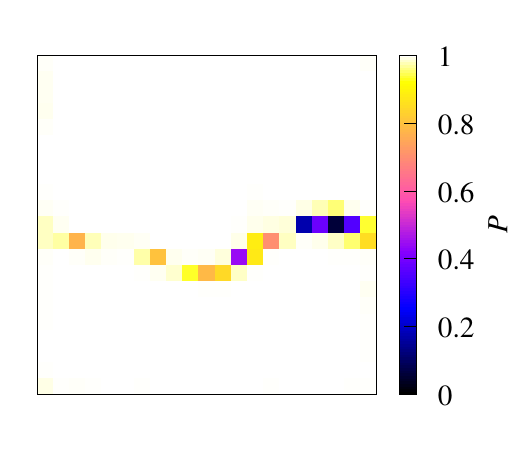}}
\end{subfigure}
%% ------------------------------------------------------------
\begin{subfigure}[b]{0.20\textwidth}
\caption{\label{fig:Lambda_a4_T0.5}}
{\includegraphics[trim={3mm 6mm 0mm 6mm},clip,height=30mm]{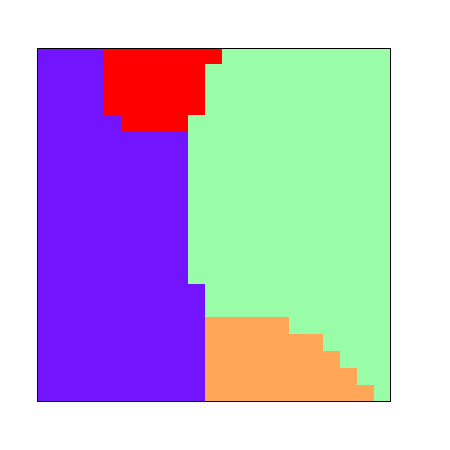}}
\end{subfigure}
%% ------------------------------------------------------------
\begin{subfigure}[b]{0.25\textwidth}
\caption{\label{fig:Prosus_a4_T0.5}}
{\includegraphics[trim={3mm 6mm 0mm 6mm},clip,height=30mm]{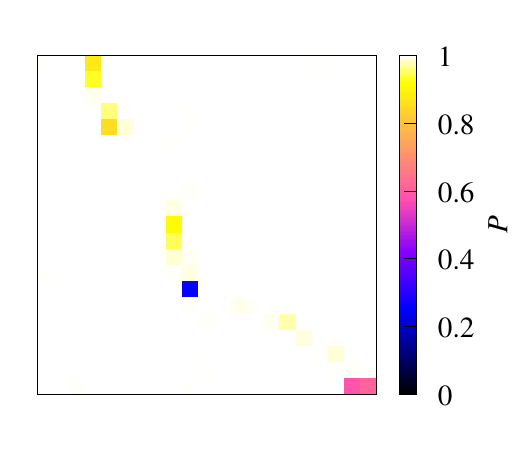}}
\end{subfigure}
\hfill %% ------------------------------------------------------------
\begin{subfigure}[b]{0.20\textwidth}
\caption{\label{fig:Lambda_a4_T1.1}}
{\includegraphics[trim={3mm 6mm 0mm 6mm},clip,height=30mm]{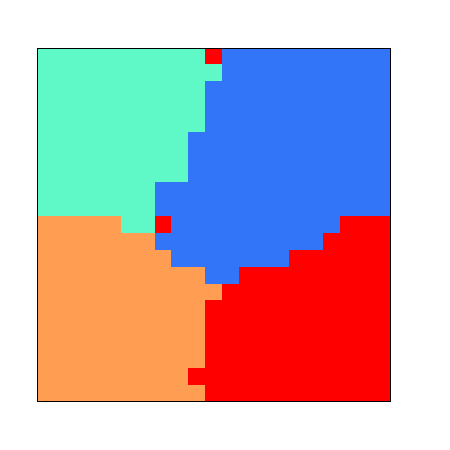}}
\end{subfigure}
%% ------------------------------------------------------------
\begin{subfigure}[b]{0.25\textwidth}
\caption{\label{fig:Prosus_a4_T1.1}}
{\includegraphics[trim={3mm 6mm 0mm 6mm},clip,height=30mm]{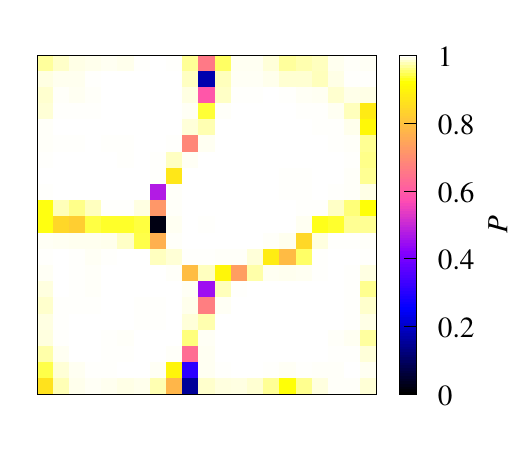}}
\end{subfigure}
%% ------------------------------------------------------------
\caption{\label{fig:Psus}Examples of maps $\lambda$ and $\mathcal P$ after $t_\textrm{max}=10^6$ for 
$\alpha=2$, $T=0.5$: \subref{fig:Lambda_a2_T0.5} $\lambda$ \subref{fig:Prosus_a2_T0.5} $\mathcal P$; 
$\alpha=2$, $T=1.1$: \subref{fig:Lambda_a2_T1.1} $\lambda$ \subref{fig:Prosus_a2_T1.1} $\mathcal P$;
$\alpha=3$, $T=0.5$: \subref{fig:Lambda_a3_T0.5} $\lambda$ \subref{fig:Prosus_a3_T0.5} $\mathcal P$; 
$\alpha=3$, $T=1.1$: \subref{fig:Lambda_a3_T1.1} $\lambda$ \subref{fig:Prosus_a3_T1.1} $\mathcal P$;
$\alpha=4$, $T=0.5$: \subref{fig:Lambda_a4_T0.5} $\lambda$ \subref{fig:Prosus_a4_T0.5} $\mathcal P$; 
$\alpha=4$, $T=1.1$: \subref{fig:Lambda_a4_T1.1} $\lambda$ \subref{fig:Prosus_a4_T1.1} $\mathcal P$;
}
\end{figure*}

In \Cref{fig:Psus} examples of maps of opinions $\lambda$ [\Cref{fig:Lambda_a2_T0.5,fig:Lambda_a2_T1.1,fig:Lambda_a3_T0.5,fig:Lambda_a3_T1.1,fig:Lambda_a4_T0.5,fig:Lambda_a4_T1.1}] and probabilities of sustaining the opinions $\mathcal P$ [\Cref{fig:Prosus_a2_T0.5,fig:Prosus_a2_T1.1,fig:Prosus_a3_T0.5,fig:Prosus_a3_T1.1,fig:Prosus_a4_T0.5,fig:Prosus_a4_T1.1}] after $t_\textrm{max}=10^6$ for various sets of parameters ($\alpha,T$) are presented.

%% ============================================================
%% ============================================================

%% #############################################################################
\section{\label{app:listing}}
%% #############################################################################

In \Cref{lst:code} the implementation of Latan\'e model rules defined by \Cref{eq:szamrej,eq:Teq0,eq:Tgt0,eq:probability_p,eq:probability_P} with distance scaling function \eqref{eq:fg} as Fortran95 code is presented.
To compile it with GNU Fortran for multi-threaded execution type\\
\fbox{\texttt{gfortran -fopenmp -O3 latane.f90}}\\
in the command line.

\begin{lstlisting}[language=Fortran,label=lst:code,caption=Source of Fortran95 code implementing Latan\'e model]
!!! Latane-Nowak-Szamrej model
!!!
!!! compile with GNU Fortran with -fopenmp
!!! for multi-threaded execution, e.g.:
!!! gfortran -O3 -fopenmp latane.f90


!!! ###########################################
module settings
!!! ###########################################
implicit none
integer, parameter :: seed = 1
logical, parameter :: randompisi=.true.
integer, parameter :: Xmax=21,Ymax=21,tmax=1e3,L2=(Xmax+1)*(Ymax+1),Run=100,Kmax=Xmax*Ymax
real*8, parameter :: alpha=2.0d0, T=0.40d0
end module settings

!!! ###########################################
module utils
!!! ###########################################
use settings
implicit none
contains

real*8 function g(x)
  real*8 :: x
  g=1.0d0+x**alpha
end function
!!! -------------------------------------------

real*8 function q(x)
  real*8 :: x
  q=x
end function
!!! -------------------------------------------

real*8 function d(x1,y1,x2,y2)
  integer :: x1,y1,x2,y2
  d=dsqrt((1.d0*x1-1.d0*x2)**2 + (1.d0*y1-1.d0*y2)**2)
end function
!!! -------------------------------------------

end module utils

!!! ###########################################
program Social_impact
!!! ###########################################
use settings
use utils
use omp_lib
implicit none
integer :: x,y,xx,yy,x1,y1,x2,y2,it,k,irun,no,counter,tau
real*8 :: r
real*8 :: sump,maxI

real*8, dimension (:,:,:,:), allocatable :: gdistance
integer, dimension (:), allocatable :: histogramno
real*8, dimension (:,:,:), allocatable :: I
real*8, dimension (:,:,:), allocatable :: prob
integer, dimension (:), allocatable :: ispresent
integer, dimension (:,:), allocatable :: lambda
real*8, dimension (:,:), allocatable :: p
real*8, dimension (:,:), allocatable :: s

allocate( gdistance(Xmax,Ymax,Xmax,Ymax) )
allocate( histogramno(L2) )

call srand(seed)
  
histogramno=0

!$  write ( *, '(a,i8)' ) &
!$    '###  OpenMP: the number of processors available = ', omp_get_num_procs ( )
!$  write ( *, '(a,i8)' ) &
!$    '###  OpenMP: the number of threads available    = ', omp_get_max_threads ( )

if(T.eq.0.0d0) then
  print '(A17)',"### deterministic"
else
  print '(A17)',"### probabilistic"
endif

if(randompisi) then
  print '(A18)',"### random s and p"
else
  print '(A18)',"### s_i=p_i=1/2   "
endif

print *,'# seed=',seed

print '(A3,7A11)','###','Xmax','Ymax','K','alpha','T','tmax','Run'
print '(A3,3I11,2F11.3,2I11)','###',Xmax,Ymax,Kmax,alpha,T,tmax,Run
print *,'######################################'
  
do x1=1,Xmax
do y1=1,Ymax
do x2=1,Xmax
do y2=1,Ymax
  gdistance(x1,y1,x2,y2)=g(d(x1,y1,x2,y2))
enddo
enddo
enddo
enddo
  
!$OMP PARALLEL SHARED(gdistance,histogramno) DEFAULT(PRIVATE)

allocate( I(Xmax,Ymax,Kmax) )
allocate( prob(Xmax,Ymax,Kmax) )
allocate( ispresent(Kmax) )
allocate( lambda(0:Xmax,0:Ymax)  )
allocate( p(Xmax,Ymax) )
allocate( s(Xmax,Ymax)  )

histogramno=0

!$OMP DO SCHEDULE(DYNAMIC)
do 777 irun=1,Run
  tau=tmax

  do x=1,Xmax
  do y=1,Ymax
    if(randompisi) then
!! random values of s and p
      s(x,y)=rand()
      p(x,y)=rand()
    else
!! homogenous values of s and p
      s(x,y)=0.5d0
      p(x,y)=0.5d0
    endif
  enddo
  enddo

  it=0
  lambda=0

  counter=1
  do x=1,Xmax
  do y=1,Ymax
!! random initial opinions
 ! lambda(x,y)=1+Kmax*rand()
!! every agent has their own opinion
     lambda(x,y)=counter
     counter=counter+1
  enddo
  enddo

!! count opinions
  ispresent=0
  do x=1,Xmax
  do y=1,Ymax
    ispresent(lambda(x,y))=1
  enddo
  enddo
  no=sum(ispresent)

!! printing system evolution
  if(irun.eq.1) then
    print *,'# irun=',irun,'it=',it,'lambda:'
    do x=1,Xmax
      print '(41I5)',(lambda(x,y),y=1,Ymax)
    enddo
    print *,"# no=",no
    print *,'######################################'
  endif

  do 88 it=1,tmax !! starting time evolution
    I=0.0d0
     
!! evaluate social impact I(x,y,\lambda) for agent at (x,y) excerted by belivers of \lambda
    do x=1,Xmax
    do y=1,Ymax
      do xx=1,Xmax
      do yy=1,Ymax
        if(lambda(x,y).eq.lambda(xx,yy)) then
          I(x,y,lambda(xx,yy))=I(x,y,lambda(xx,yy))+q(s(xx,yy))/gdistance(x,y,xx,yy)
        else
          I(x,y,lambda(xx,yy))=I(x,y,lambda(xx,yy))+q(p(xx,yy))/gdistance(x,y,xx,yy)
        endif
      enddo
      enddo
    enddo
    enddo

    do x=1,Xmax
    do y=1,Ymax
    do k=1,Kmax
      I(x,y,k)=4.0d0*I(x,y,k)
    enddo
    enddo
    enddo
  
!! evaluate a probability that agent at (x,y) will take opinion \lambda
    do x=1,Xmax
    do y=1,Ymax
      sump=0.0d0
      do k=1,Kmax
        prob(x,y,k)=0.d0 !! new
        if(I(x,y,k).gt.0.0d0) prob(x,y,k)=dexp(I(x,y,k)/T) !! new if
        sump=sump+prob(x,y,k)
      enddo
      do k=1,Kmax
        prob(x,y,k)=prob(x,y,k)/sump
      enddo
    enddo
    enddo

!! set (new) opinion of an agent at (x,y)
    if(T.eq.0.0d0) then !! deterministic case   
      do x=1,Xmax
      do y=1,Ymax
        maxI=I(x,y,1)
        lambda(x,y)=1
        do k=2,Kmax
          if(I(x,y,k).gt.maxI) then
            maxI=I(x,y,k)
            lambda(x,y)=k
          endif
        enddo
      enddo
      enddo
    else !! probabilistic case
      do x=1,Xmax
      do y=1,Ymax
        r=rand()
        sump=0.0d0
        do k=1,Kmax
          sump=sump+prob(x,y,k)
          if(r.lt.sump) goto 666
        enddo
666     lambda(x,y)=k
      enddo
      enddo
    endif

!! count opinions
    ispresent=0
    do x=1,Xmax
    do y=1,Ymax
      ispresent(lambda(x,y))=1
    enddo
    enddo
    no=sum(ispresent)
    if(no.eq.1) goto 22
    
88 enddo !! ending time evolution
 
22 continue

!$omp critical(hist)
histogramno(no)=histogramno(no)+1
!$omp end critical(hist)

!! printing system evolution
  if(irun.eq.1) then
    print *,'# irun=',irun,'it=',it,'lambda:'
    do x=1,Xmax
      print '(21I5)',(lambda(x,y),y=1,Ymax)
    enddo
    print *,"# no=",no
    print *,'######################################'
  endif

777 enddo
!$OMP END DO

deallocate( I )
deallocate( prob )
deallocate( ispresent )
deallocate( lambda )
deallocate( p )
deallocate( s )

!$OMP END PARALLEL

print '(A2,A4,4A11)',"#","K","T","alpha"
print '(A2,I4,4F11.3)',"#",Kmax,T,alpha
     
print *, "# histogram of the final number of opinions n_o^u"
do k=1,L2
  if(histogramno(k).gt.0) print *,k,histogramno(k)
enddo
  
deallocate( gdistance )
deallocate( histogramno )
  
end program Social_impact
\end{lstlisting}

\end{document}